\documentclass[twocolumn,tighten]{aastex63}

\newcommand{\ltsima}{$\; \buildrel < \over \sim \;$}
\newcommand{\simlt}{\lower.5ex\hbox{\ltsima}}

\newcommand{\Nii}{[N~{\sc II}]}
\newcommand{\Sii}{[S~{\sc ii}]}
\newcommand{\Oiii}{[O~{\sc iii}]}
\newcommand{\Oi}{[O~{\sc i}]}
\def\arcdeg{\hbox{$^\circ$}}
\def\arcmin{\hbox{$^\prime$}}
\def\arcsec{\hbox{$^{\prime\prime}$}}

\turnoffedit{}

\usepackage{amsmath}
\usepackage{graphicx}
\usepackage{hyperref}

\begin{document}

\title{Observational Constraints on Cool Gas Clouds in M82's Starburst-Driven Outflow}

\shorttitle{M82 Cool Clouds}
\shortauthors{Lopez et al.}

\correspondingauthor{Sebastian Lopez}
\email{lopez.764@osu.edu}

\author[0000-0002-2644-0077]{Sebastian Lopez} 
\affil{Department of Astronomy, The Ohio State University, 140 W. 18th Ave., Columbus, OH 43210, USA}
\affil{Center for Cosmology and AstroParticle Physics, The Ohio State University, 191 W. Woodruff Ave., Columbus, OH 43210, USA}

\author[0000-0002-1790-3148]{Laura A. Lopez}
\affil{Department of Astronomy, The Ohio State University, 140 W. 18th Ave., Columbus, OH 43210, USA}
\affil{Center for Cosmology and AstroParticle Physics, The Ohio State University, 191 W. Woodruff Ave., Columbus, OH 43210, USA}

\author[0000-0003-2377-9574]{Todd A. Thompson}
\affil{Department of Astronomy, The Ohio State University, 140 W. 18th Ave., Columbus, OH 43210, USA}
\affil{Center for Cosmology and AstroParticle Physics, The Ohio State University, 191 W. Woodruff Ave., Columbus, OH 43210, USA}
\affil{Department of Physics, The Ohio State University, 191 W. Woodruff Ave., Columbus, OH 43210, USA}

\author[0000-0002-2545-1700]{Adam K. Leroy}
\affil{Department of Astronomy, The Ohio State University, 140 W. 18th Ave., Columbus, OH 43210, USA}
\affil{Center for Cosmology and AstroParticle Physics, The Ohio State University, 191 W. Woodruff Ave., Columbus, OH 43210, USA}

\author[0000-0002-5480-5686]{Alberto D. Bolatto} 
\affil{Department of Astronomy, University of Maryland, College Park, MD 20742, USA}
\affil{Joint Space-Science Institute, University of Maryland, College Park, MD 20742, USA}

\begin{abstract}

Star formation feedback can drive large-scale, multi-phase galactic outflows. The dynamical and thermodynamical interaction between the hot and cooler phases is a prime focus of both observational and theoretical work. Here, we analyze H$\alpha$-emitting structures in the extraplanar wind of the nearby starburst M82. We use high-resolution, narrow-band, observations from the Hubble Legacy Archive \citep{Mutchler2007}. Our analysis constrains the morphology, number density, and column density of the structures. We highlight conspicuous arc-like structures that differ significantly from the linear cometary clouds that emerge from galactic wind simulations and discuss their possible origins, such as bow shocks or instabilities driven by cosmic rays. The most prominent structures range in size from $\sim24 -110$\,pc. Using the H$\alpha$ brightness and assumptions about the depth of the emitting structures, we estimate number densities of $\sim1-23$\,cm$^{-3}$ \edit1{assuming a unity volume filling factor},  which are lower than previous constraints from spectroscopic nebular line studies. The derived column densities, $\sim10^{20}-10^{21}$\,cm$^{-2}$, along the path of the outflow are above theoretical thresholds for cool cloud survival in a hot supersonic background, but small enough that the structures could be accelerated by the hot wind momentum. Using diffuse X-ray emission maps from $\textit{Chandra}$, we also find that even on small ($\sim100$~pc) scales, the H$\alpha$ ``leads" the X-rays, a behavior long noted in the literature on kiloparsec scales. This behavior, along with previous observational studies of ionization in the wind, may signal that shock ionization is responsible for the H$\alpha$ emission we observe.


\end{abstract}


\section{Introduction}


Rapidly star-forming galaxies (``starbursts") produce multi-phase outflows driven by stellar feedback \citep{Veilleux2005,Rubin2014,Veilleux2020,Thompson2024}. Galactic winds affect galaxy evolution by ejecting metals from the host and enriching the surrounding circumgalactic medium (CGM) and intergalactic medium (IGM) \citep{Tumlinson2017}. These winds are multiphase and appear across the electromagnetic spectrum. These various phases are the hot $10^{6-7}$~K X-ray-emitting phase \citep{Strickland2004a,Strickland2004b}, the warm $10^4$~K ionized gas emitting in the UV/optical \citep{Westmoquette2009b,Westmoquette2009a}, the neutral $10^4$~K or cooler gas in absorption \citep{Rupke2005} and emission \citep{Martini2018}, the cold gas emitting in the radio/mm \citep{Bolatto2013,Leroy2015}, and the dust emitting in the infrared (IR) \citep{Engelbracht2006,Bolatto2024} that all contribute to galactic wind phenomenology. The interplay between these phases is an active area of research.

A major open question is the origin, acceleration, and evolution of the cooler ($<10^6$~K) galactic wind phases. The classical \cite{CC85} (CC85) model predicts a very hot ($>10^6$~K) central starburst medium energized by stellar feedback in the form of energy and mass injection from supernovae and massive star winds. The model predicts the thermodynamical and dynamical properties of a superheated flow that expands adiabatically outside the starburst core \citep{Strickland2009}. While the CC85 model describes the hot phase, it leaves open the question of how the observed high-velocity cool gas is formed and accelerated \citep{Thompson2024}. The classical picture is that cool ISM clouds are ram pressure accelerated by the hot phase (e.g., \citealt{Scannapieco2015,Zhang2017}). However, recent work suggests that there is a complicated thermodynamic and dynamical interplay between the hot and cooler phases, where energy and momentum are exchanged as cool clouds interact with the background hot wind material \citep{gronke_growth_2018,Gronke2020,gronke_survival_2022,fielding_structure_2022}. Another possibility is that the hot X-ray-emitting material might become radiative if it is sufficiently mass-loaded, producing cool material via bulk cooling \citep{Wang1995, Silich2004, Thompson2016}, as might be observed in NGC~253 \citep{Thompson2016, Lopez2023}.

\begin{figure}
    \centering
    \includegraphics[width=\columnwidth]{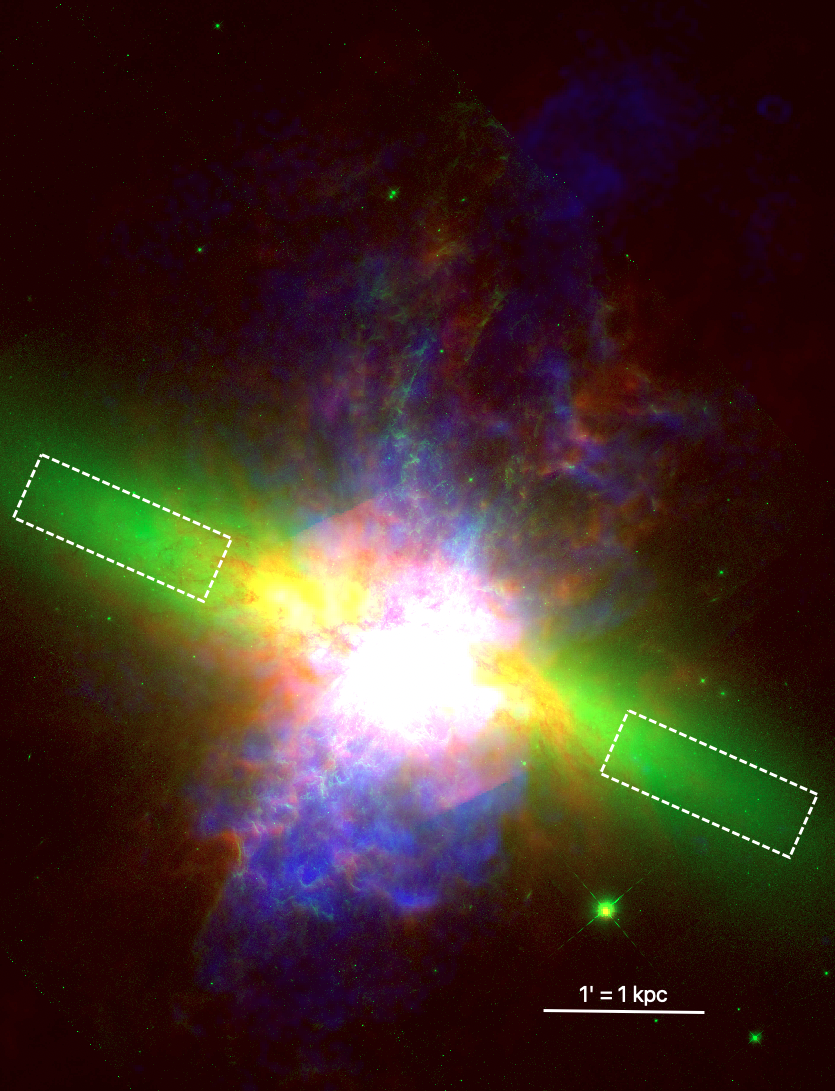}
    \caption{Three color image of M82, where blue is broad-band (0.5-7 keV) {\it Chandra} X-rays \citep{Lopez2020}, green is the HST F658N image \citep{Mutchler2007}, and red is Spitzer 8 $\mu$m \citep{Kennicutt2003,Engelbracht2006} infrared emission. The white dashed boxes are the areas used for the continuum subtraction in Section~\ref{sec:dataproc}. At the 3.6~Mpc away, 1\arcmin{} is about 1~kpc as shown in the scale bar. In this image and all subsequent ones, north is up and east is left.}
    \label{fig:3color}
\end{figure}

Previous work has long noted and analyzed the relationship between cool clouds and galactic winds. In NGC~1569, \cite{Heckman1995} found H$\alpha$ filaments to be co-spatial with X-ray spurs found in ROSAT data at a resolution of 8.6\arcsec. The spectra of the H$\alpha$ filaments showed high-velocity and low-velocity components, where the origin of the latter was hypothesized to be cool clouds flowing outward. In M82, \cite{Lehnert1999} also analyzed the co-spatial X-ray and H$\alpha$ emission in the wind. They found that the X-ray properties can be explained as the result of shock heating as the hot wind interacts with cooler surrounding gas. These shocks can also produce H$\alpha$ emission in clouds carried by the wind. \cite{Westmoquette2007,Westmoquette2009b,Westmoquette2009a} came to similar conclusions about the observational signatures of cool clouds. Using HST spectra and GMOS-IFU data, they found a consistent low-velocity, broad-line component that they asserted was from entrained clouds. They concluded that the smaller clouds have their outer layers stripped off by the hot gas and that this material is subsequently incorporated into the wind.

As the prototypical inclined starburst galaxy, the nearby (3.6~Mpc; \cite{freedman94,gerke11}), nearly edge-on (80$^{\circ}$; \citealt{mckeith95}) galaxy M82 has served as a key target in this field, and recent work has also attempted to measure the properties of individual entrained clouds in the M82 wind directly. \cite{Krieger2021} used NOEMA CO data to measure the properties of CO emission up to 3~kpc away from M82’s disk. They identified and characterized 764 molecular clouds in M82’s wind, finding sizes of $\sim$50~pc comparable to the $30$~pc resolution of their data. In contrast, \cite{Xu2023} analyzed nebular line emission from ionized gas and found clouds to have radii below one parsec \citep[see also][]{Westmoquette2009b}. They derived their constraints by measuring density using the \Sii~$\lambda\lambda6717,\:6731$ doublet, calculating a volume, and inferring cloud size from derived filling factors. 
\cite{Fisher2024} characterized the structures in the inner 870~pc $\times$ 870~pc of M82's wind using 3.3~$\mu$m JWST infrared data that trace polycyclic aromatic hydrocarbons (PAHs), which follow the cool gas. They found cloud structures, with widths of about 10~pc and heights spanning up to 150~pc, resemble the cometary structures seen in both cloud wind-tunnel simulations \citep{gronke_growth_2018,Gronke2020,gronke_survival_2022,Abruzzo2023,Tan2024,Villares2024} and in global simulations \citep{Cooper2008,Schneider2020}. 

These studies have used dust and molecular emission and optical spectroscopy to characterize the properties of clouds in the wind of M82, but there has so far been no systematic study of the morphology of the cool ($\sim 10^4$~K) H$\alpha$-emitting clouds seen in the M82 outflow at the high resolution \edit1{achieved} by HST. 
In this paper, we analyze the high resolution H$\alpha$ (F658N) images of the M82 outflow acquired by the Hubble Legacy Team\footnote{Based on observations made with the NASA/ESA Hubble Space Telescope, and obtained from the Hubble Legacy Archive, which is a collaboration between the Space Telescope Science Institute (STScI/NASA), the Space Telescope European Coordinating Facility (ST-ECF/ESA) and the Canadian Astronomy Data Centre (CADC/NRC/CSA).} and published in \cite{Mutchler2007}. This image has high enough spatial resolution ($\sim1$~pc) that we can visually identify the entrained cloud structures and to measure their physical properties for comparison with simulations and observations of other phases in the M82 wind \citep[e.g., the hot phase visible in X-ray emission analyzed by][]{Lopez2020}. This H$\alpha$ emission traces gas at a temperature of approximately $10^4$~K, which is a typical value for both the temperature floor and cloud temperature in galactic wind simulations \citep{Schneider2020}.

We structure the paper as follows: In Section~\ref{sec:methods}, we detail the data used and its processing, discuss the selection of clouds, and explain how we calculate various cloud properties, such as their sizes, emission measures, densities, column densities, and masses. In Section~\ref{sec:results}, we present our results, detailing new constraints on the arc-like morphology and properties of the structures observed. In Section~\ref{sec:discuss}, we compare our findings to both simulations that model the cloud-wind interactions and past observational studies that characterize cool clouds in galactic winds. We also discuss the ionization mechanisms that produce the H$\alpha$ emission and the possible origins of the clouds. We adopt a distance of 3.6~Mpc to M82 throughout this paper, such that 1\arcmin\ $\approx$ 1~kpc.

\section{Methods}\label{sec:methods}

\subsection{Archival Data Used and Data Processing}
\label{sec:dataproc}

In Figure~\ref{fig:3color}, we present a three-color image of M82 showing the data used in this analysis. The hot ($>10^6$~K), X-ray emitting phase is shown in blue where the data acquired was from the \textit{Chandra} X-ray Observatory as detailed in \cite{Lopez2020} (doi:\dataset[10.25574/cdc.320]{https://doi.org/10.25574/cdc.320} ). Green shows \textit{Hubble} Space Telescope (HST) observations using the F658N filter (\citealt{Mutchler2007}; doi:\dataset[10.17909/T9GW2P]{https://doi.org/10.17909/T9GW2P}), which captures mostly starlight in the disk and mostly H$\alpha$ emission from warm ($\sim10^4$K) ionized gas in the outflow. \textit{Spitzer} imaging of 8~$\mu$m (IRAC4) infrared emission is shown in red (\citealt{Kennicutt2003}; doi:\dataset[10.26131/IRSA424]{https://doi.org/10.26131/IRSA424}). \textit{Spitzer}'s 8$\mu$m filter captures mostly emission from the 7.7$\mu$m PAH feature with some contribution from an underlying dust and stellar continuum.


Our analysis focuses on the H$\alpha$ emission captured by the F658N filter. These observations were obtained as part of HST proposal 10776 (PI: Matt Mountain) and are described in \citet{Mutchler2007}. The observations also include the F555W and F814W filters, and we obtained the high level science products associated with this project from the Hubble Legacy Archive (HLA). 

We use the images provided with the original \citet{Mutchler2007} release,  so we do not benefit from, e.g., astrometric alignment to \textit{Gaia} or any improvements to the HST pipeline. To account for this, we updated the astrometry of the HLA images to match more recent F814W observations of M82. These were taken in 2019 as part of HST proposal 15645 (PI: David Sand;  doi: \dataset[10.17909/5bm4-pf48]{http://dx.doi.org/10.17909/5bm4-pf48}), and we obtained them from from the Mikulski Archive for Space Telescopes (MAST). We identified matching point sources in the HLA \citet{Mutchler2007} and new MAST F814W images and updated the HLA astrometry using the \textsc{ASTROPY} function \textsc{fit\_wcs\_from\_points}. The new astrometric solution was then applied to the F555W and F658N images from the HLA.

The F658N filter captures emission from stars, H$\alpha$ nebular emission, and \Nii{} nebular emission. Therefore stellar continuum subtraction and an estimate of the \Nii{}/H$\alpha$ ratio are required for our analysis. To isolate the nebular emission in the F658N filter, we used the adjacent F555W and F814W HST filters. 

Before accounting for the continuum, we background subtracted, smoothed, and converted the units for each image. To set the background level, we defined a 0.5\arcmin{} $\times$ 0.5\arcmin{} region 3.1~kpc from the disk, calculated the average intensity in this region, and subtracted it from the entire image. We also converted the units of each of the images from electrons s$^{-1}$ pixels$^{-1}$ to intensity units of erg s$^{-1}$ cm$^{-2}$ sr$^{-1}$ \r{A}$^{-1}$. This conversion used the {\sc photflam} factor from the FITS header, which has units of erg cm$^{-2}$ \r{A}$^{-1}$ electrons$^{-1}$, and the pixel scale to convert pixels to steradians. For the F658N we also used the filter FWHM of 72~\r{A} to remove the \r{A} unit, leaving that image with units of erg s$^{-1}$ cm$^{-2}$ sr$^{-1}$. To avoid differences among the point-spread-functions (PSF) at the different wavelengths of the images, we also convolved the data to 0.1\arcsec{} resolution, which is coarser than the lowest resolution image (F814W). 

Once the background subtraction and unit conversions were completed, we combined the F555W and F814W images to predict the stellar continuum in the F658N filter. To do this, we linearly interpolated between the filter midpoints, weighting by wavelength, i.e.,
\begin{equation}
\label{eq:stellarcont}
I^{\rm F658N}_{\rm stellar} = \frac{(814-658) I^{\rm F814W} + (658-555) I^{\rm F555W}}{814-555}
\end{equation}
where $I^{F814W}$ refers to the intensity in the F814W filter. $I^{\rm F658N}_{\rm stellar}$ is the predicted intensity of the stellar continuum in units of erg~s$^{-1}$~cm$^{-2}$~sr$^{-1}$~\r{A}$^{-1}$. We apply Eq. \ref{eq:stellarcont} pixel by pixel and so create a predicted stellar continuum image.

\begin{figure*}
    \centering
    \includegraphics[width=\textwidth]{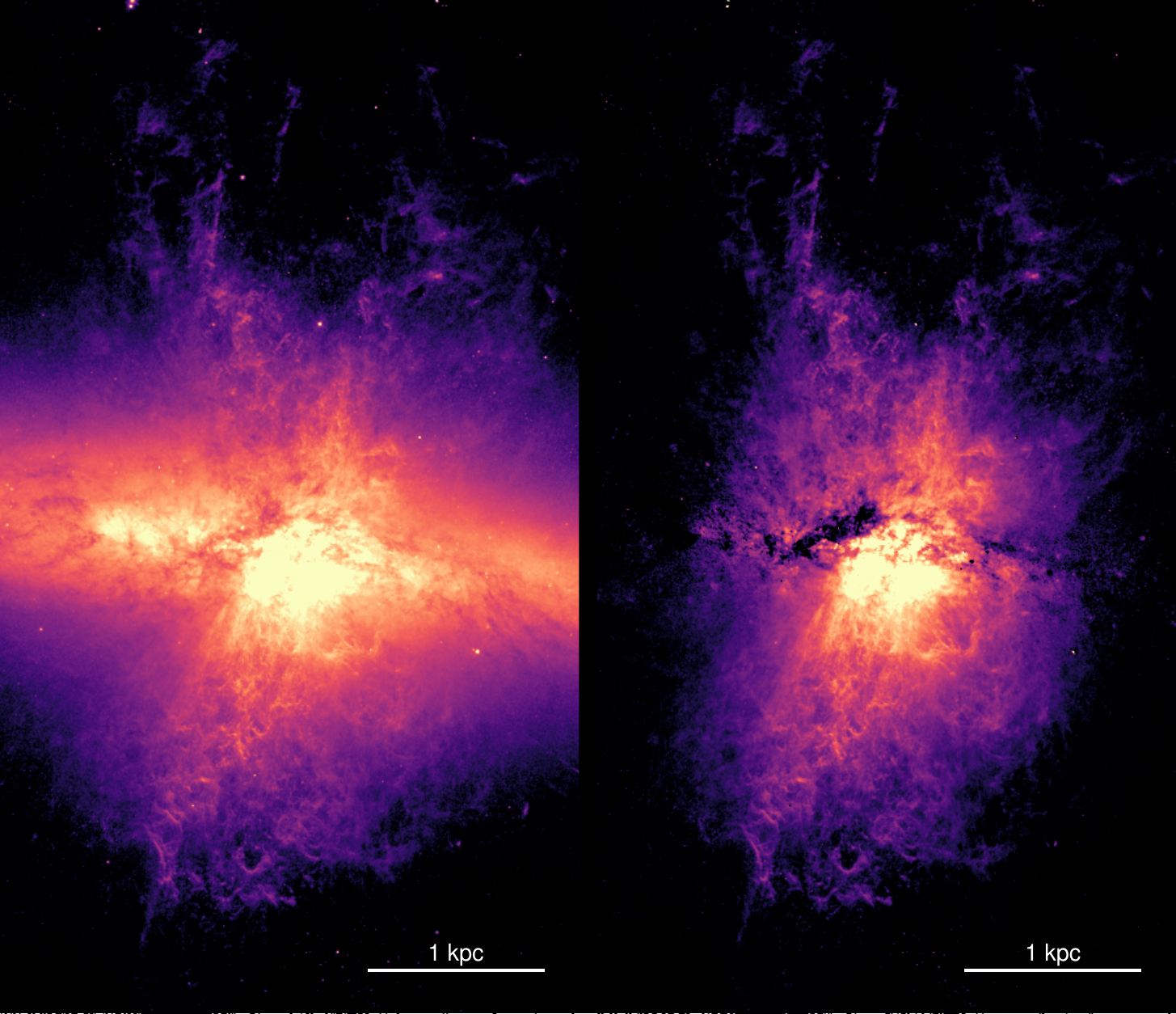}
    \caption{\textit{Left}: HST F658N image. The image includes  starlight, which is most evident in the stellar disk but also affects the starburst and outflow regions. \textit{Right}: F658N after subtracting the stellar continuum subtraction using the F555W and F814W images from \cite{Mutchler2007}. The stellar disk has been successfully removed from the image and the images shows only H$\alpha$+\Nii{} nebular emission. }
    \label{fig:cont_sub}
\end{figure*}

We rescaled this predicted $I^{\rm F658N}_{\rm stellar}$ image to match the F658N image in regions where starlight dominates the actual F658N image. To accomplish this, we defined two  1.5\arcmin{} $\times$ 0.5\arcmin{} regions on the western and eastern sides of the disk that appear dominated by starlight and shown them as the white boxes in Fig. \ref{fig:3color}. The scaling factor was then calculated by taking the ratio of the F658N image over the continuum image in those regions. We found a median scaling factor of $\sim74$. We scaled the continuum image by that factor and then subtracted it from the F658N image. In Figure~\ref{fig:cont_sub} we show the convolved F658N \edit1{image} before continuum subtraction on the left and the post-continuum subtraction version on the right. The stellar disk is no longer apparent in the continuum subtracted image, and the H$\alpha$+\Nii{} in the wind remains.

To test the robustness of our continuum subtraction, we compared the HST H$\alpha$+\Nii{} and continuum images to those from the Local Volume Legacy (LVL) survey \citep{Dale2009}. We first convolved the HST H$\alpha$+\Nii{} and continuum images to have a PSF of $\sim 1.7''$ to match PSF of the LVL data, which we determined from measuring the brightness profiles of several stars in the LVL continuum image. 

\begin{figure}
    \centering
    \includegraphics[width=0.47\textwidth]{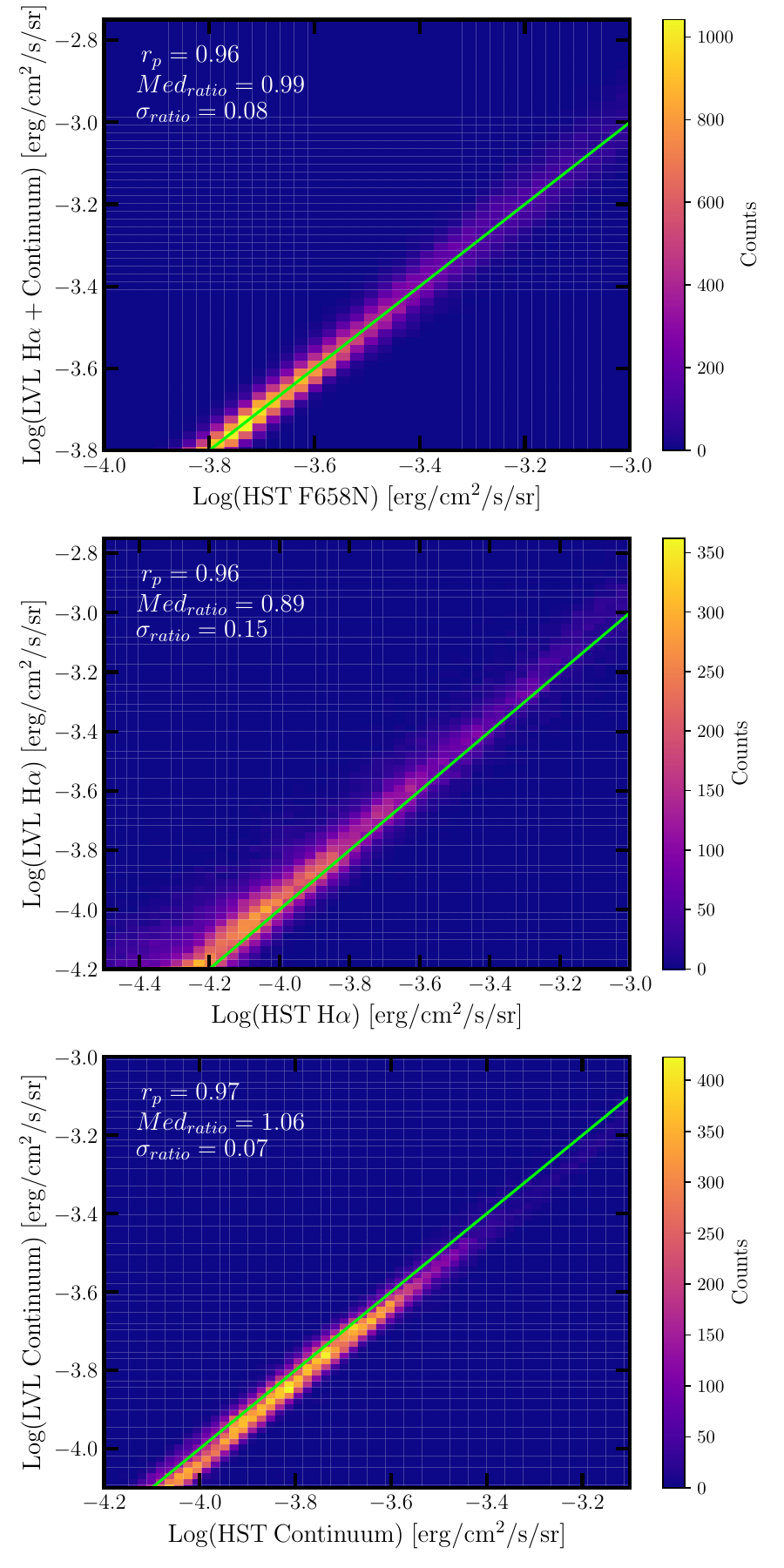}
    \caption{Comparison of the HST F658N data to LVL imaging. Each panel compares the intensity of pixels in the two data sets after matching the resolution and astrometry of the HST data to the LVL image.  \textit{Top}: the line plus continuum images. \textit{Middle}: The continuum subtracted H$\alpha$+\Nii{} only images. \textit{Bottom}: the stellar continuum estimate and our stellar continuum estimate created from the HST F555W and F814W images. In each panel, the one-to-one line is shown in green and the color scale shows the density pixels in the intensity-intensity space. In the top left we report the Pearson correlation coefficient ($r_{p}$) relating pixels in the two images, the median ratio between the data sets, and the scatter in the ratio. All panels show overall good agreement, demonstrating the validity of our continuum subtraction routine and suggesting an overall $\lesssim 10{-}15\%$ systematic uncertainty in our H$\alpha$+\Nii{} images.}
    \label{fig:lvlcomp}
\end{figure}

Figure~\ref{fig:lvlcomp} \edit1{compares} the line-only, continuum, and combined LVL and HST images. The top panel compares the line plus continuum LVL data to our HST F658N data before any continuum subtraction. The two agree well, with a Pearson corelation coefficient ($r_{\rm p}$) of 0.96, a median ratio between the images of 0.99, and a scatter of 8\%. The middle panel of Figure~\ref{fig:lvlcomp} shows the correlation between our continuum subtracted H$\alpha$+\Nii{ }F658N image and its LVL counterpart. We find overall good agreement, though slightly worse than the top panel, with $r_{\rm p}=0.96$, a median ratio of 0.89, and 15\% scatter. The bottom panel compares the LVL and HST continuum images, which have $r_{\rm p}=0.97$, a median ratio of 1.06, and 7\% scatter. The quantitative similarities between our images gives confidence in our continuum subtraction and validates the calibration of the images, implying systematic uncertainties of $\sim 10{-}15\%$.

We identified point sources using the {\sc photutils} function DAOStarfinder. We then masked these sources and filled them in by interpolating the surrounding pixels. \edit1{We then corrected for the effects of Milky Way extinction by multiplying the image by $10^{A_R/2.5}$ with $A_R=0.339$~mag. This is the foreground extinction in the Landolt R-band filter for this line of sight from \citep{Schlafly2011}. We chose the R-band because it contains the F658N filter bandpass. }

Finally, we account for \Nii{} emission in the F658N filter. We calculated the \Nii{}/H$\alpha$ ratio across the outflow using data from the pathfinder Dragonfly Spectral Line Mapper (pDSLM) \citep{Abraham2014,Lokhorst2020}. \cite{Pasha2021} and \cite{Lokhorst2022} published pDSLM maps for the M81$-$M82 system.

The Dragonfly pDSLM dataset consists of separate, continuum-subtracted H$\alpha$ and \Nii{} images, which we further processed by masking point sources and subtracting a local background. We calculated this average background level from versions of the images with M81 and M82 masked and subtracted it from the H$\alpha$ and \Nii{} images. Then we converted from units of counts to \edit1{$\mathrm{erg/s/cm^{2}}$} following the procedure in \cite{Lokhorst2022}, where fluxes were calibrated by comparing \ion{H}{2} region measurements in M81 to literature values (this yielded $\mathrm{log_{10}(}F\;\mathrm{[erg\;s^{-1}\;cm^{-2}]}) = a\;\mathrm{log_{10}(}F\;\mathrm{[counts]}) + b$, with $a = 1.00 \pm 0.01$ and $b = -0.86 \pm 0.04$). To reduce noise, we smoothed the background-subtracted images with a 5\arcsec{}$\times$5\arcsec{} Gaussian kernel. We then cropped the images to a 10\arcmin{}$\times$10\arcmin{} region centered on M82 and retained only pixels with a signal-to-noise ratio (SNR) of 10 or higher. The noise was calculated within a 3\arcmin{}$\times$3\arcmin{} area located away from the galaxy and picked to be free of point sources and diffuse gas emission. These high-SNR \Nii{} and H$\alpha$ images were then divided to derive a map of the \Nii{}/H$\alpha$ ratio across the outflow. We reprojected this ratio map back onto the HST F658N astrometric grid, multiplied with the continuum subtracted H$\alpha$+\Nii{} image, and isolated the H$\alpha$ emission using the expression

\begin{equation}
\label{eq:niisub}
I_{\rm H\alpha}^{\rm F658N} = I^{\rm{F658N}}_{\rm H\alpha + \Nii{}} / (1 +\rm{\Nii{}/H\alpha})
\end{equation}

After correcting for extinction and contamination by \Nii{}, we converted our intensity image into units into emission measure (EM) using
\begin{equation}
    \mathrm{EM} = \frac{4\pi\:I_{\rm H\alpha}^{\rm F658N}}{\alpha_{\rm{eff, H\alpha}}(T)\:h\nu},
\end{equation}
where $I^{\rm 658N}_{\rm H\alpha}$ is the H$\alpha$ line intensity in each pixel, and $\alpha_{\rm{eff,H\alpha}}(T)=1.17\times10^{-13}\;\rm{cm^3\;s^{-1}}$ is the effective recombination coefficient for H$\alpha$ at $T\simeq10^4$~K \citep{Draine2011}. We assume Case B recombination.

\subsection{Cloud Selection}

\begin{figure*}
    \centering
    \includegraphics[width=0.88\textwidth]{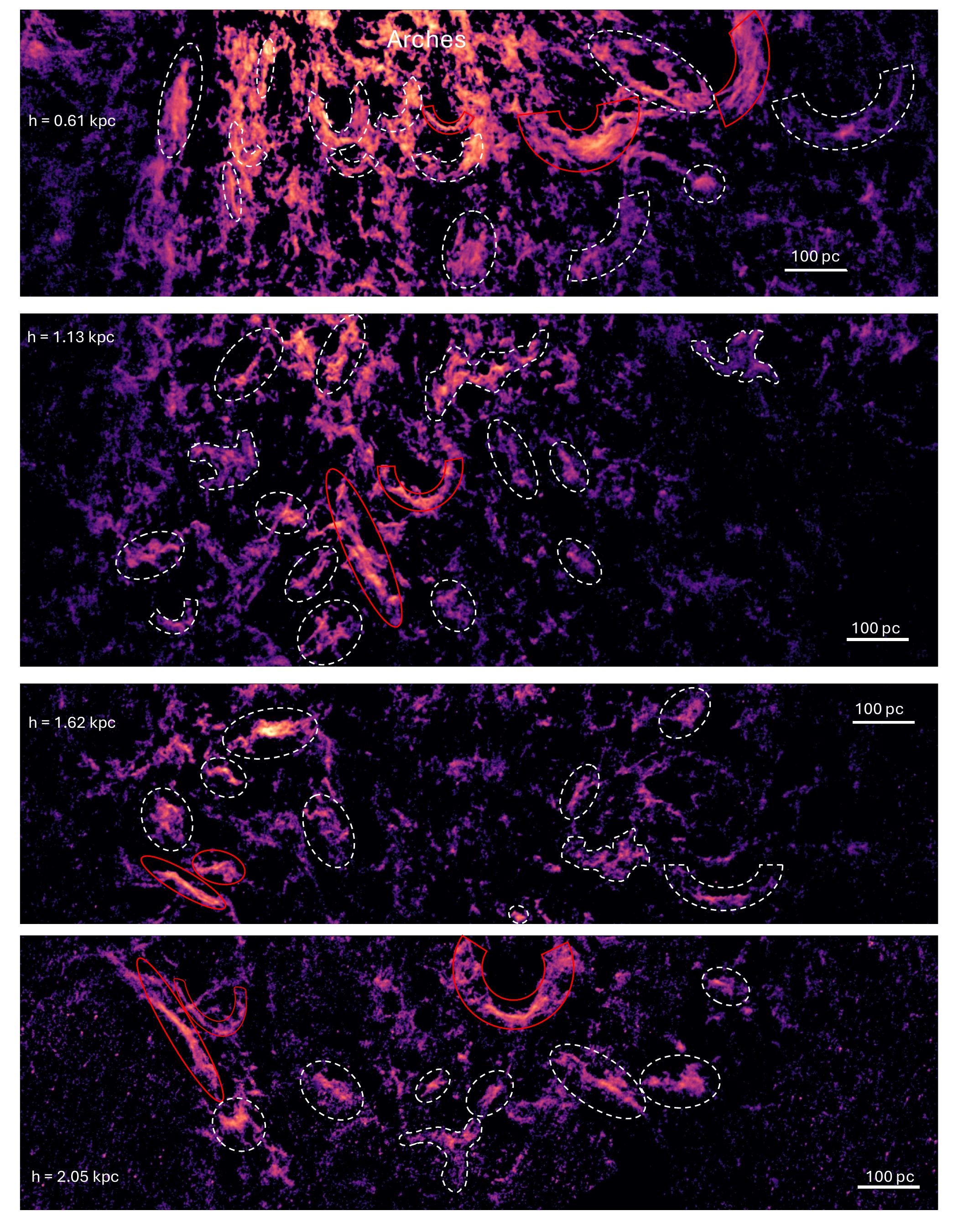}
    \caption{Slices of the continuum-subtracted HST H$\alpha$ image of M82's southern outflow from the Hubble Legacy Archive \citep{Mutchler2007} shown in Figure~\ref{fig:cont_sub}. An unsharp mask filter (detailed in Section~\ref{sec:methods}) has been applied to enhance the filamentary structure of the outflow and facilitate selecting cloud structures. The red contours are the clouds analyzed in this paper. The white dashed contours show some of the many other structures present in the outflow ranging from arcs and ellipsoids to highly complex geometries. The vertical distance from the disk and a scale bar of 100~pc is shown in each slice. }
    \label{fig:south_zoomin_usm}
\end{figure*}

\begin{figure*}
    \centering
    \includegraphics[width=0.71\textwidth]{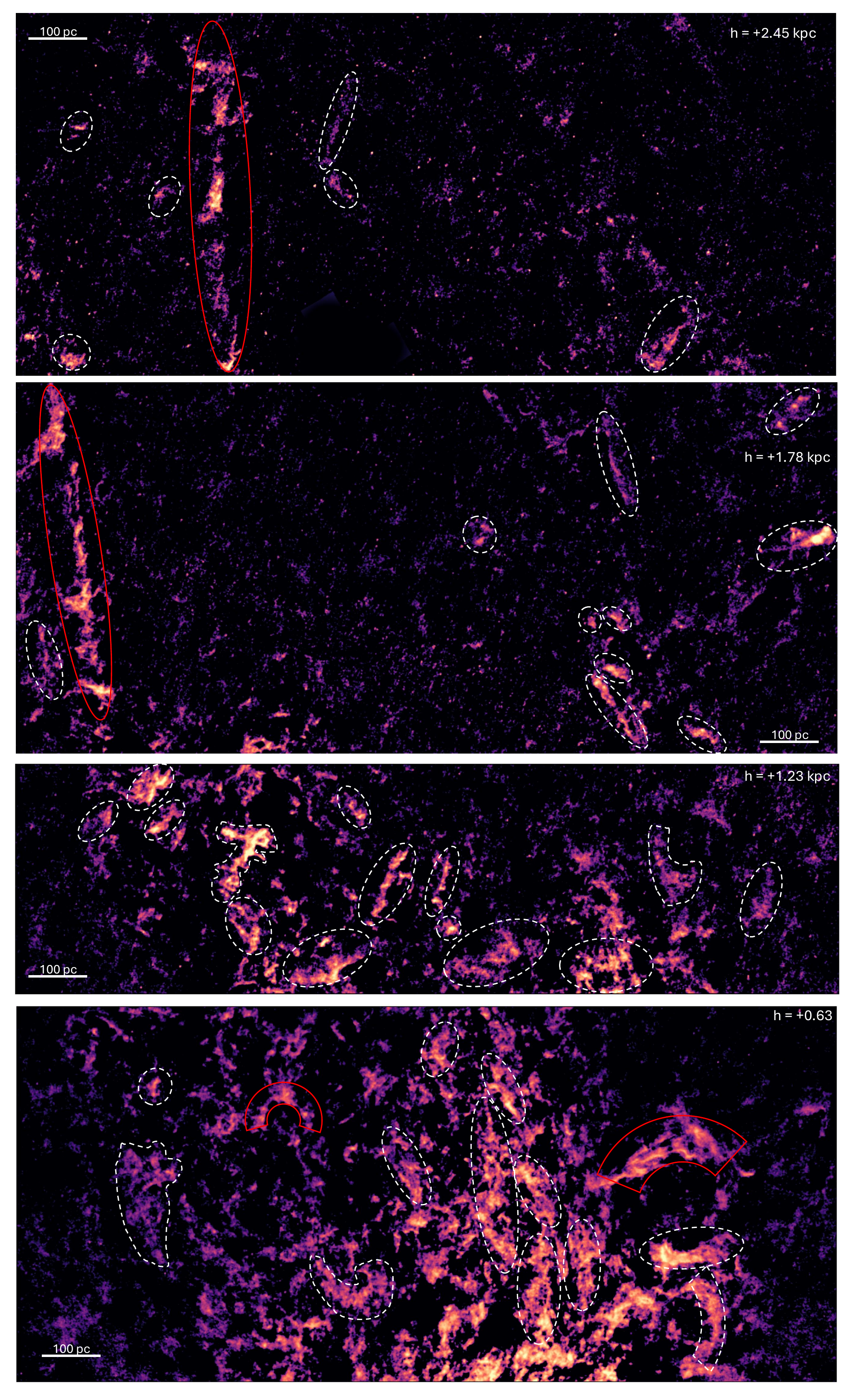}
    \caption{Same as Figure~\ref{fig:south_zoomin_usm} but for the northern outflow of M82. Red contours show the clouds studied in this paper, and white dashed contours show other examples of clouds found in the wind. The northern outflow in particular has several elongated like structures and few arcs unlike the south. The vertical distance from the disk and a scale bar of 100~pc is shown in each slice.}
    \label{fig:north_zoomin_usm}
\end{figure*}

\edit1{Quantitative approaches have often been taken to characterize filamentary structures in the ISM such as in the works of \cite{Rosolowsky2008} \cite{Makarenko2015}. However, in this paper we take a more qualitative approach by selecting the structures through visual inspection.} To further enhance the filamentary structure in the wind and to help select spatially separated clouds, we created an unsharp mask version of the processed H$\alpha$ image and show it in Figure~\ref{fig:south_zoomin_usm} for the southern outflow and in Figure~\ref{fig:north_zoomin_usm} for the northern outflow. The unsharp mask was created by convolving the image with a Gaussian kernel that has a FWHM of 5\arcsec. The smoothed image was then subtracted from the original image (right panel of Figure~\ref{fig:cont_sub}), multiplied by a factor of 10, and the result was added back to the original image. The masked image was used only for identifying unique structures, while the processed unmasked image was used for the remainder of the analysis.

\begin{figure*}
    \centering
    \includegraphics[width=0.84\textwidth]{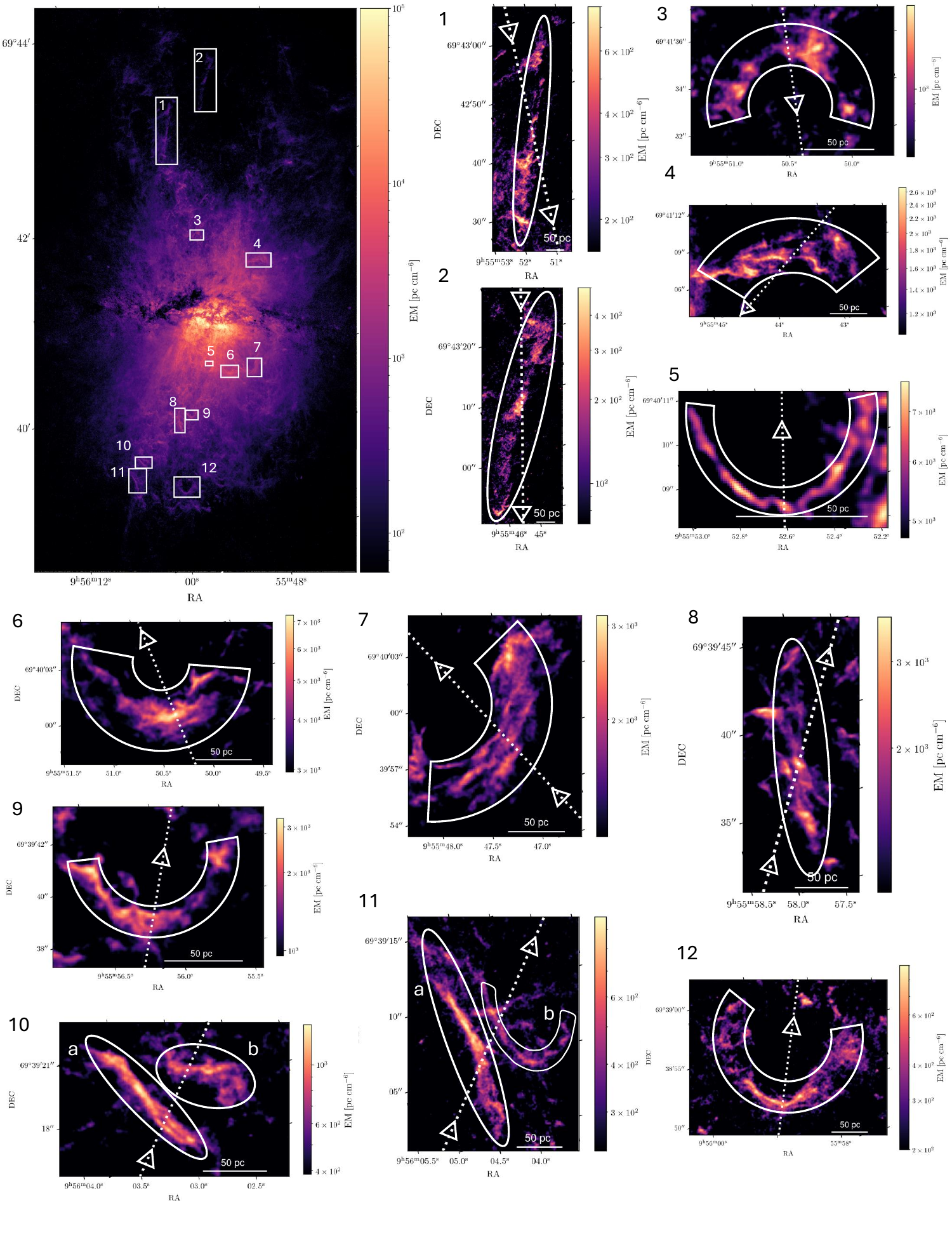}
    \caption{The top left panel is an emission measure map of M82. White boxes are the viewing windows where individual cloud structures are selected for further analysis. The other panels are zoom-in images of the viewing windows overplotted in the left hand side image. Once again the color map is the emission measure. Regions 10 and 11 are further divided due to multiple identified structures. The overlaid contours are the assumed geometries for all 14 clouds. Regions 1, 2, 3, 8, 10a, 10b, and 11a are all approximated as ellipsoids. Regions 4, 5, 6, 7, 9, and 12 are approximated by circular arcs and region 11b is an elliptical arc. The white arrows point toward the starburst core and the dotted line is the axis used to measure the path length of the cloud the hot wind traverses. These geometries are used to produce the results in Table~\ref{table:values}.}
    \label{fig:cloudatlas}
\end{figure*}

Based on this unsharp mask, we created 12 cutouts of different structures throughout the northern and southern outflows, which we show in Figure~\ref{fig:cloudatlas}. 
These cutouts were centered on a representative sample of the structures visually identified in Figures~\ref{fig:south_zoomin_usm} and \ref{fig:north_zoomin_usm} which we highlight with the red contours. Each cutout highlights one structure, except 10 and 11, which have two visibly identified structures, respectively. For each of the clouds, we assumed various geometric shapes to characterize their morphology. The shapes were chosen through visual inspection and created using SAOImage DS9. In Figure~\ref{fig:cloudatlas}, we show cutouts of the clouds along with their assigned geometry. Regions 1, 2, 3, 8, 10a, 10b, and 11a are all approximated as ellipses. Regions 4, 5, 6, 7, 9, and 12 are approximated by circular arcs, and region 11b is approximated as an elliptical arc.

We also assumed a three-dimensional geometry associated with each type of shape. This allows us to infer a line-of-sight depth for each cloud and to estimate the volume and volume density of gas. Using these assumed geometries, we measured the major and minor axes of the elliptical clouds, the outer and inner radii of the circular arc clouds, and both of the former set of values for the elliptical arc cloud 11b.  For the elliptical structures like regions 1 and 2, we assumed the depth into the image is the minor axis. For the circular arcs like regions 5 and 12, we assumed the depth is the difference between the outer and inner radii (like a semi-circular wedge), and for the elliptical arc in region 11b we assume the depth is the average of the differences between the major and minor axes. To compare to simulations that quote a single characteristic cloud radius $r_{\rm{cl}}$ \citep{Tan2024}, we use our assumed three dimensional geometry to derived $r_{\rm{cl}}$ by equating the calculated volume of each cloud to that of a sphere and solving for the radius as $r_{\rm{cl}} = (3V/4\pi)^{1/3}$.


We note that there are several possible geometric approximations for characterizing the cloud morphology. For instance, regions 1 and 2 could also be represented as cylinders rather than ellipsoids, and the arcs may instead be semi-hemispherical shells, yielding a similar appearance. While it is challenging to determine the true geometry of these clouds due to projection effects from the edge-on outflow, approximating their morphology is essential for constraining their physical properties. We demonstrate the effects of varying geometry for cloud 12 in Section~\ref{sec:cloud_phys_props}.

The assumed geometries also provide constraints on cloud properties, including number density $n_{\rm e}$, column density $N_{\rm H}$, and cloud size $r_{\rm{cl}}$, which can be compared with recent cloud survival simulations. For simplicity, we assume all the observed gas is hydrogen. If the medium is optically-thin, the observed specific intensity $I$ of H$\alpha$ emission along the line of sight is given by
\begin{equation}
    \frac{dI}{ds} =j 
    \Longrightarrow  I = n_{\rm e}^2\alpha_{\rm{eff,H\alpha}}(T)\frac{h\nu}{4\pi}f\Delta s, 
    \label{eq:int}
\end{equation}
where $j$ is the emissivity and $s$ is the dimension into the sky. Assuming a depth $\Delta s$ of a given \edit1{structure, and a volume filling factor $f$}, we can use this expression for the intensity to obtain an estimate of the number density of the emitting gas:
\begin{equation}
    n_{\rm e} \simeq \left(\frac{4\pi I}{\alpha_{\rm{eff,H\alpha}}(T)\;f\Delta s\;h\nu}\right)^{1/2}.
    \label{eq:ne1}
\end{equation}
Scaling for parameters typical of the structures we identify in the H$\alpha$ maps,
\begin{equation}
    n_{\rm e} \simeq 10.9\;\mathrm{cm^{-3}}\left(\frac{I}{10^{-5}\:\mathrm{\frac{erg}{s\;cm^2\;sr}}}\right)^{1/2}\left(\frac{\rm pc}{\Delta s}\right)^{1/2}\left(\frac{1}{f}\right)^{1/2}.
    \label{eq:ne2}
\end{equation}
In applying these formulae to the structures we identify, we assume that $I$ is the average intensity within the assigned cloud shape. $\Delta s$ is the assumed depth into the image, which depends on the structure geometry as described above. We note that this type of calculation is biased towards higher densities because of the $n^2$ dependence of the emission measure. \edit1{Further, in reporting the quantities derived below for the structures we identify, we assume $f=1$ for simplicity, but explicitly include the scaling with the filling factor. We note that the gas densities reported are then lower limits.  This is an important uncertainty we return to in Section~\ref{sec:comp_obs}, where we compare with the spectral line  study of \cite{Xu2023}.}

Another parameter of interest for comparing with simulation results is the projected gas column density $N_{\rm H}$ parallel to the wind direction \citep{gronke_growth_2018,Gronke2020,gronke_survival_2022}, which emanates from the starburst nucleus:
\begin{equation}
    N_{\rm H} = n_{\rm e}f^{1/2}\Delta l,
    \label{eq:col}
\end{equation}
where $\Delta l$ is the cloud structure dimension parallel to the wind \edit1{and $f$ is once again the volume filling factor}. As shown by the dotted white line in Figure~\ref{fig:cloudatlas}, \edit1{we measure the path length the hot wind transverses through the center of the each region.}


\section{Results}\label{sec:results}

\subsection{Observed Morphology}

Figure~\ref{fig:3color} reveals a network of filamentary structures traced by the optical and infrared emission that are correlated with the hot phase. The filaments are even clearer in Figures~\ref{fig:south_zoomin_usm} and \ref{fig:north_zoomin_usm}, which show zoom-ins of the southern and northern outflows that are further exaggerated with an unsharp mask filter. These filaments exhibit a complex morphology that differs from the cometary structures predicted in simulations \citep{Cooper2008,gronke_growth_2018}. The structures whose properties we derive in this paper are highlighted by the red contours, and in the white contours are other possible structures that are identifiable throughout the wind.

We also find differences between the hemispheres. The southern outflow contains prominent arcs spanning from 0.5 to almost 3~kpc away. Most notably in Figure~\ref{fig:south_zoomin_usm} is the collection of arcs near the starburst base that we label as the ``Arches". These arcs appear to be aligned with each other along the major axis of the galaxy, and their inner parts appear hollow. We discuss what could be creating these structures in Section~\ref{sec:discuss}. The northern outflow, on the other hand, contains fewer prominent arcs (clouds 3 and 4). Instead, in the north, there are several elongated line-like structures that we characterize as ellipsoids and that appear more cometary. The quantity of structures in the north also appear more sparse than in the south. Both outflows have structures that are not easily characterized as arcs or ellipsoids; they instead have more complicated morphologies, shown by the polygons in Figures~\ref{fig:south_zoomin_usm} and \ref{fig:north_zoomin_usm}.

\begin{deluxetable*}{cccccccccccccccc} \rotate
\tablecolumns{15}
\tablewidth{0pt} \tablecaption{Cloud Properties \label{table:values}\tablenotemark{a}\tablenotemark{b}}
\tablehead{\colhead{Cloud} & \colhead{Shape\tablenotemark{c}} &\colhead{Distance}  & \colhead{$R$}& \colhead{$r$}  &\colhead{$\Delta s$} &\colhead{$\Delta l$} & \colhead{$r_{\rm cl}$} &\colhead{$I$} &\colhead{$n_{\rm e}$} &\colhead{$N_{\rm H}$}  & \colhead{$\chi$} & \colhead{$t_{\rm cc}$} & \colhead{$t_{\rm cool, mix}$} & \colhead{$M_{\rm cl}$}  \\
\colhead{} & \colhead{} & \colhead{(kpc)} & \colhead{(pc)} & \colhead{(pc)} & \colhead{(pc)} & \colhead{(pc)} & \colhead{(pc)} & \colhead{($10^{-5}\;\mathrm{\frac{erg}{s\;cm^2\;sr}}$)} &\colhead{($\mathrm{cm^{-3}}$)} & \colhead{($\mathrm{ 10^{20} \;cm^{-2}}$)} & \colhead{} & \colhead{(Myr)} & \colhead{($10^{-3}\;\mathrm{Myr}$)} & \colhead{($10^4\;\mathrm{M_\odot})$} \\
\colhead{(1)} & \colhead{(2)} & \colhead{(3)} & \colhead{(4)} & \colhead{(5)} & \colhead{(6)} & \colhead{(7)} & \colhead{(8)} & \colhead{(9)} & \colhead{(10)} & \colhead{(11)} & \colhead{(12)} & \colhead{(13)} & \colhead{(14)} & \colhead{(15)} }  
\startdata
1 & E & 2.08 & 338 & 40.2 & 80.3 & 236 & 81.7 & $1.86 \pm 0.3$ & $1.6 \pm 0.1$ & $11.9 \pm 1.0$ & $81.5 \pm 6.6$ & $0.36 \pm 0.01$ & $6.47 \pm 0.53$ & $9.2 \pm 0.8$ \\ 
2 & E & 2.57 & 338 & 62.9 & 126& 418 & 110 & $0.70 \pm 0.3$ & $0.8 \pm 0.1$ & $10.3 \pm 1.8$ & $27.5 \pm 4.9$ & $0.28 \pm 0.03$ & $13.2 \pm 2.37$ & $11 \pm 2.0$ \\ 
3 & CA & 0.88 & 61.0 & 30.5 & 30.5 & 30.4 & 33.4 & $6.38 \pm 0.5$ & $4.9 \pm 0.2$ & $4.61 \pm 0.2$ & $140 \pm 6.0$ & $0.19 \pm 0.00$ & $2.15 \pm 0.09$ & $1.9 \pm 0.1$ \\ 
4 & CA & 0.76 & 161 & 83.0 & 77.9 & 100 & 68.5 & $10.1 \pm 1.0$ & $3.7 \pm 0.2$ & $11.9 \pm 0.6$ & $110 \pm 5.2$ & $0.35 \pm 0.01$ & $2.73 \pm 0.13$ & $13 \pm 0.6$ \\ 
5 & CA & 0.48 & 36.7 & 26.2 & 10.5 & 10.5 & 14.4 & $46.8 \pm 1.7$ & $23 \pm 0.4$ & $7.3 \pm 0.1$ & $236 \pm 4.3$ & $0.11 \pm 0.00$ & $0.47 \pm 0.01$ & $0.7 \pm 0.0$ \\ 
6 & CA & 0.60 & 80.7 & 26.2 & 54.6 & 55.4 & 49.7 & $29.4 \pm 1.3$ & $7.9 \pm 0.2$ & $13.5 \pm 0.3$ & $82.0 \pm 1.9$ & $0.22 \pm 0.00$ & $1.34 \pm 0.03$ & $10 \pm 0.2$ \\ 
7 & CA & 0.69 & 105 & 51.2 & 54.2 & 57.5 & 51.0 & $12.5 \pm 1.4$ & $5.2 \pm 0.3$ & $9.1 \pm 0.5$ & $147\pm 7.9$ & $0.30 \pm 0.01$ & $2.05 \pm 0.11$ & $7.0 \pm 0.4$ \\ 
8 & E & 1.16 & 114 & 25.4 & 50.8 & 122 & 41.8 & $11.3 \pm 0.4$ & $5.1 \pm 0.1$ & $19 \pm 0.4$ & $163 \pm 3.1$ & $0.26 \pm 0.00$ & $2.09\pm 0.04$ & $3.8 \pm 0.1$ \\ 
9 & CA & 1.05 & 56.1 & 35.7 & 20.4 & 20.6 & 24.5 & $12.6 \pm 0.5$ & $8.4 \pm 0.15$ & $5.3 \pm 0.1$ & $240 \pm 4.4$ & $0.19 \pm 0.00$ & $1.25 \pm 0.02$ & $1.3 \pm 0.0$ \\ 
10a & E & 1.73 & 67.6 & 14.6 & 29.2 & 29.5 & 24.3 & $5.4 \pm 0.3$ & $4.6 \pm 0.14$ & $4.2 \pm 0.1$ & $231 \pm 7.2$ & $0.18 \pm 0.00$ & $2.29 \pm 0.07$ & $0.7 \pm 0.0$ \\ 
10b & E & 1.69 & 40.5 & 24.2 & 48.4 & 44.6 & 28.7 & $4.6 \pm 0.3$ & $3.3 \pm 0.1$ & $4.6 \pm 0.2$ & $165 \pm 5.9$ & $0.18 \pm 0.00$ & $3.19 \pm 0.11$ & $0.8 \pm 0.0$ \\ 
11a & E & 1.94 & 130 & 24.1 & 48.2 & 65.6 & 42.3 & $2.7 \pm 0.3$ & $2.6 \pm 0.2$ & $5.2 \pm 0.3$ & $128 \pm 7.3$ & $0.23 \pm 0.01$ & $4.13 \pm 0.23$ & $2.0 \pm 0.1$ \\ 
11b & EA\tablenotemark{d} & 1.87 & 72.7 & 46.2 & 26.5 & 22.7 & 27.8 & $2.7 \pm 0.3$ & $3.4 \pm 0.2$ & $2.4 \pm 0.1$ & $170 \pm 9.9$ & $0.18 \pm 0.01$ & $3.10 \pm 0.18$ & $0.8 \pm 0.0$ \\ 
12\tablenotemark{e} & CA & 1.82 & 106 & 62.0 & 44.4 & 44.2 & 54.0 & $2.7 \pm 0.3$ & $2.6 \pm 0.2$ & $3.6 \pm 0.2$ & $132 \pm 8.3$ & $0.30 \pm 0.01$ & $4.00 \pm 0.25$ & $4.3 \pm 0.3$
\enddata
\tablenotetext{\edit1{a}}{\edit1{All values assume a filling factor of 1 (see Equations~\ref{eq:ne2} and \ref{eq:col}).}}
\tablenotetext{b}{Column 1 is the assigned cloud number. Column 2 is the assumed geometric shape for the cloud. Column 3 is the distance from the starburst center. Column 4 is the semi-major axis of the ellipse or the outer radius of the circular arc and Column 5 is the same but for the semi-minor axis and inner radius of the arc. Column 6 is the assumed distance into the image to calculate $n_e$ in Equation~\ref{eq:ne2}. Column 7 is the distance across the cloud that the hot wind traverses used to calculate $N_H$ in Equation~\ref{eq:col}. Column 8 is the cloud radius calculated by equating the volume of the cloud to that of a sphere and solving for radius. Column 9 is the average intensity within the cloud shapes defined in Figure~\ref{fig:cloudatlas}. Column 10 is the number density of the cloud and column 11 is the column density of the cloud parallel to the wind. Column 12 is the density contrast ratio between the derived H$\alpha$ emitting cloud $\rm{n_e}$ and the surrounding hot wind values from \cite{Lopez2020}. Column 13 is the cloud crushing time assuming a wind velocity of $v_w = 2000~\rm{km\;s^{-1}}$ and column 14 is the cooling time of the cloud mixing layer assuming $T_{mix}=10^{5.5}$~K. Column 15 is the mass of the cloud assuming the H$\alpha$ emitting gas fills the whole cloud volume.}
\tablenotetext{c}{The shape abbreviations are E for ellipse, CA for circular arc, and EA for elliptical arc.}
\tablenotetext{d}{Values for $R$ and $r$ are the average semi-major and semi-minor axis. The semi-major axis of inner ellipse is 57.9~pc and the semi-minor axis is 35.8~pc. The semi-major axis of the outer ellipse is 85.4~pc and the semi-minor axis is 52.8~pc} 
\tablenotemark{e}{For Cloud 12 we consider alternative geometries shown in Figure~\ref{fig:cloud_dissect} and discuss their effects on the derived values in Section~\ref{sec:cloud_phys_props}.}
\end{deluxetable*}

\subsection{Cloud Physical Properties}

\label{sec:cloud_phys_props}
\begin{figure*}
    \centering
    \includegraphics[width=0.91\textwidth]{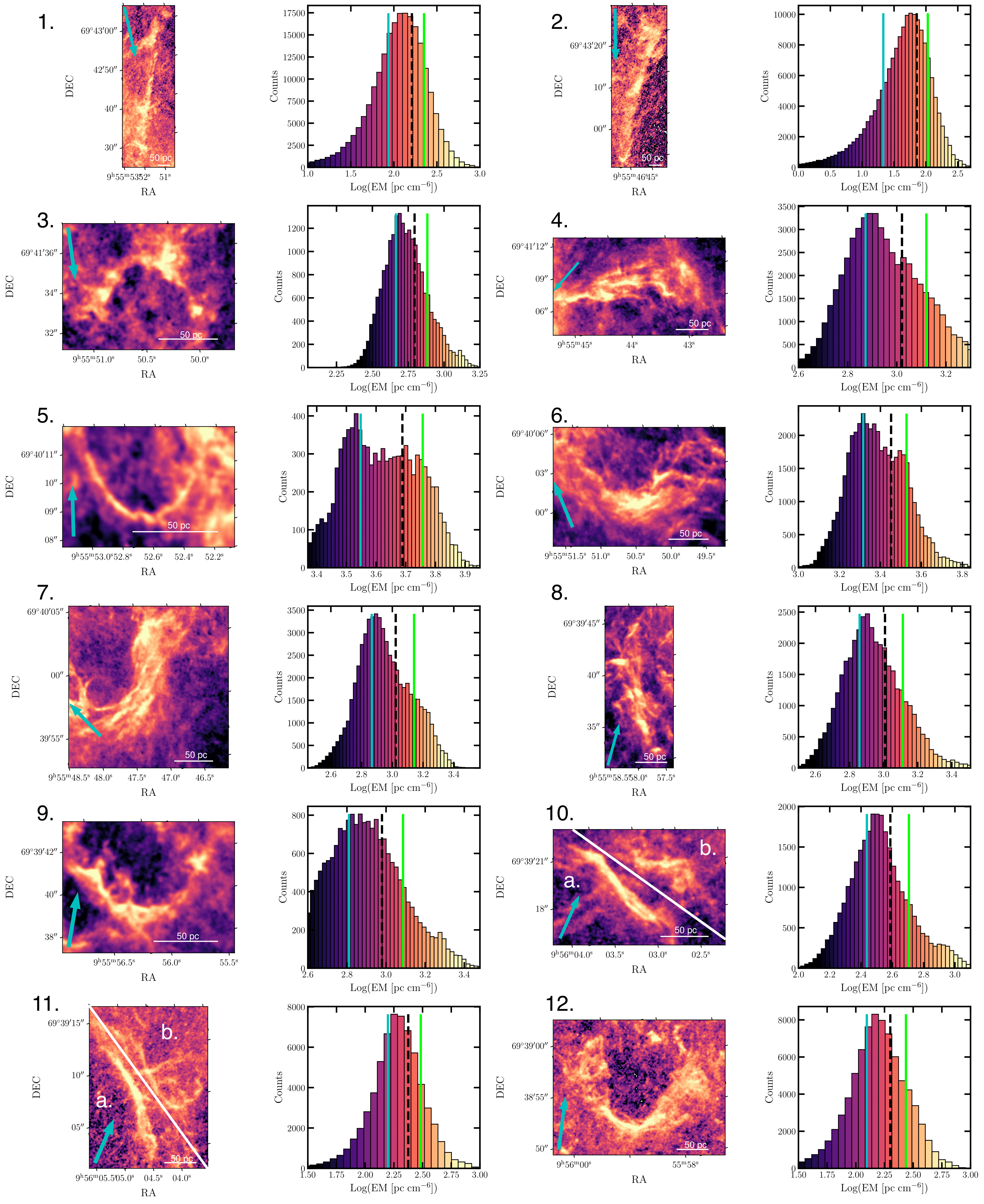}
    \caption{Images of clouds and histograms showing the distribution of emission measure values from each region. The colormap of the histograms is set to that of the images highlighting that the structures are present in regions of high emission measure. The cyan arrows point toward the starburst core. The cyan line is the median of the diffuse gas values around the cloud, and the green line is the median of the cloud values. The black dashed line shows the division between the diffuse gas and cloud material. The scale bar shows a size of 50 pc. The first four regions are shown in decreasing distance from the starburst core. The latter 8 regions are shown in increasing distance from the starburst core. As structures are closer to the starburst core, the emission measure distribution spans higher values.}
    \label{fig:em_hists}
\end{figure*}

In Figure~\ref{fig:em_hists}, we present the emission measure (EM) distribution for each of the 12 box regions shown in Figure~\ref{fig:cloudatlas}. The emission measure values range from $1$ to $10^4\;\mathrm{pc\;cm^{-6}}$, with the regions farther from the disk having lower values than those nearer to the starburst center, where the emission measure peaks. The clouds account for the highest values in each histogram, as evident in the distributions, which are colored by the range of emission measure values. This illustrates the contrast between the density of the gas that forms the clouds and the more diffuse background. 

To better quantify the contrast, we overplot the median lines for the diffuse gas values in cyan and the emission associated with the cloud in green. The division between the cloud and the surrounding diffuse gas is found to be best characterized visually by the 68\% percentile brightness level, where anything brighter than this threshold is considered to be the cloud and anything dimmer is instead associated with the diffuse gas. We show how the clouds look with the 68\% percentile cutoff in Figure~\ref{fig:cloudatlas}; we find it to highlight solely the cloud emission well. 
This threshold is denoted by a vertical black dashed in the emission measure histogram. With this approach we find that the clouds have 1.6 $-$ 5 times higher emission measure than the surrounding diffuse emission.

For each cloud, we compile several physical properties in Table~\ref{table:values}. As described in Section~\ref{sec:methods}, these include the gas number density $n_{\rm e}$ and column density of the clouds $N_{\rm H}$ (eqs.~\ref{eq:ne2} \& \ref{eq:col}) \edit1{and all assume $f=1$}. 
We reiterate that the projected column densities listed in Table~\ref{table:values} are for the path parallel to the wind direction (dashed white lines in Figure~\ref{fig:cloudatlas}), making them relevant for comparison to cloud survival criteria discussed in Section~\ref{sec:comp_sims}. 
The number densities range from $n_{\rm e}$ = 0.80 $-$ 23~$\mathrm{cm^{-3}}$ and decrease with distance from the starburst. \edit1{Given the dependence of equations \ref{eq:ne2} and \ref{eq:col} on the volume filling factor $f$, the gas densities $n_e$ and column densities $N_{\rm H}$ are lower  and upper limits, respectively.}

Our measured column densities are as low $N_{\rm H} = 2.4\times10^{20}\mathrm{cm^{-2}}$, as large as $N_{\rm H} = 1.2\times10^{21}\mathrm{cm^{-2}}$, and have no trend with distance from the starburst, at least for the small number of structures we consider. Cloud mass $M_{\rm cl}$ is also constrained using the number densities $n_{\rm e}$ and volumes of the clouds, using their assumed geometries. For simplicity, we assume the H$\alpha$-emitting gas fills the entire cloud; however, it is possible that the H$\alpha$ emission is confined to the surface of the cloud, with cooler material inside, which we discuss in the next section. We find cloud masses between $M_{\rm cl} \approx 6.9\times10^3$ and $1.3\times10^5\;\mathrm{M_\odot}$ \edit1{for $f=1$}.

Another common parameter in wind-cloud simulations is the density contrast $\chi$ between the denser cool gas clouds and the more tenuous hot wind background. To calculate $\chi$ for the cloud structures in Table \ref{table:values}, we use our derived number densities $n_{\rm e}$ together with with those from \cite{Lopez2020}, who calculated the hot gas properties of the wind as a function of distance from the disk using \textit{Chandra} X-ray data. For each of the spatial bins used by \cite{Lopez2020}, we use the X-ray flux-weighted hot gas number density to calculate $\chi$ for each cloud structure. \edit1{For $f=1$,} we find density contrasts of the order of $\chi \approx 10^{2}$ (listed in Column 12 of Table~\ref{table:values}), similar to those assumed in cloud-crushing simulations \citep{gronke_growth_2018,Tan2024}. 

Using our derived $\chi$ values, we calculate a cloud-crushing time, $t_{\rm cc} = \chi^{1/2} r_{\rm cl}/v_{\rm w}$, where $v_{\rm w}$ is the hot wind velocity. Since the dynamics of the hot gas have not been directly measured from X-ray spectroscopy, constraints on the hot wind velocity come from modeling of the thermal content of the very hot gas in the central starburst. Theoretical models fitted to X-ray data \citep{Strickland2009,Nguyen2021} predict the wind speed to be $1000-2000$~km~s$^{-1}$. However, recent observational work by \cite{Boettcher2024} using XMM-Newton RGS spectra indicates that the soft X-rays are produced by a hot wind with velocity of at least $\sim2000$~km~s$^{-1}$. Using the values of $\chi$ and $r_{\rm cl}$ derived for each structure, and assuming a fiducial value of $v_{\rm w}=2000$~km~s$^{-1}$, we calculate the cloud-crushing times to be in the range from $t_{\rm cc} = 0.11 - 0.36$~Myr, as shown in column 13 in Table~\ref{table:values}. We also derive the cloud mixing layer cooling times using the formula $t_{\rm cool, mix}=3k_B T_{\rm mix}/n_e\Lambda$, where we assume $T_{\rm mix}=10^{5.5}$~K and the cooling function is $\Lambda=10^{-21.4}$ erg cm$^3$ s$^{-1}$. The cooling time of this layer is the one relevant for cloud survival since that is the material that can cool rapidly and potentially make the cloud grow at the expense of the hot wind material. We find the cooling times to be $t_{\rm cool, mix} \approx 10^{-3}-10^{-2}$~Myr, as shown in column 14 in Table~\ref{table:values}, one to two orders of magnitude lower than the cloud crushing times. We discuss the implications of this comparison in Section~\ref{sec:comp_sims}.

\begin{figure*}
    \centering
    \includegraphics[width=\textwidth]{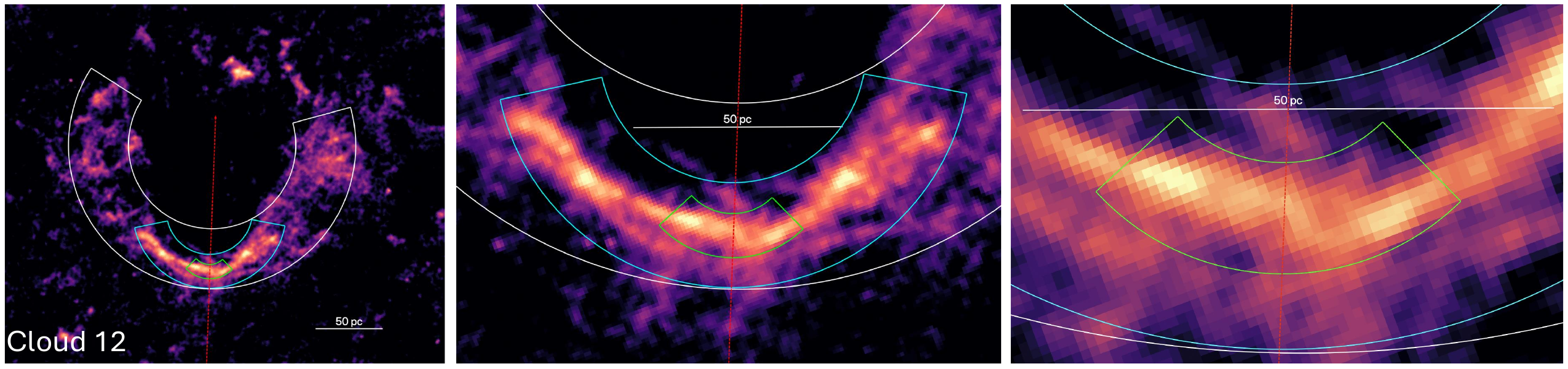}
    \caption{\textit{Left:} Different arc geometries considered for cloud 12. The white arc is the nominal shape whose results are shown in Table~\ref{table:values}. \textit{Middle:} The medium size arc is focused on the brightest arc center with an outer radius of 56.7~pc and an inner radius of 31.9~pc. \textit{Right:} The smallest arc is focused on one of the brightest sections of the arc in the middle panel. The outer radius is 23.9~pc and the inner radius is 13.4~pc. The red dotted line points to the center of the galaxy and is where the length of the structure is measured to calculate $N_{\rm{H}}$.}
    \label{fig:cloud_dissect}
\end{figure*}

As shown in Figure~\ref{fig:cloud_dissect} for cloud 12, we also consider different arc sizes to determine their effects on the properties listed in Table~\ref{table:values}. The white contour (left panel) is the nominal cloud geometry, whose results are in Table~\ref{table:values}. The cyan contour (left and middle panels) focuses on the smaller bright central arc. Assuming this geometry, we find $n_{\rm e}\simeq4.2\:\rm{cm^{-3}}$, $N_{\rm H} \simeq 3.2\times10^{20}\;\rm{cm^{-2}}$, $\chi\simeq209$, and $t_{\rm cc} \simeq 0.19$~Myr. These values differ from the derived values for the fiducial geometry assumed in Table 1 by factors of 1.7, 0.94, 1.7, and 0.63, respectively. The green contour (right panel) focuses on one of the brightest sections of the central arc. Assuming this geometry for this particular subregion, we find $n_{\rm e}\simeq7.7\:\rm{cm^{-3}}$, $N_{\rm H} \simeq 2.5\times10^{20}\;\rm{cm^{-2}}$, $\chi\simeq384$, and $t_{\rm cc} \simeq 0.09$~Myr. The subregion differs from the original cloud by factors of 3.1, 0.73, 3.1, and 0.30, respectively. \edit1{For each geometry and for each reported property, we assume $f=1$.} These variations in the derived density and column density within the individual substructures of cloud 12 highlight the fact that each structure we analyze (Table~\ref{table:values}) has a range of cooling times, density contrasts with the surrounding hot flow, and cloud crushing times.

\subsection{Caveats on the measured cloud properties}

Along with the derived cloud properties in Table~\ref{table:values}, we also list their corresponding uncertainties. However, we note that these errors are solely the statistical uncertainties of the intensity measurements propagated for each quantity. To derive these uncertainties, we began by defining emission-free regions in the original HLA F658N image and calculated the mean absolute deviation (MAD). This value serves as the baseline error for each pixel in the image. When conducting the continuum subtraction, we adopted an additional 15\% uncertainty on the intensity to account for the error in our continuum subtraction. This uncertainty was defined by the scatter shown in the middle panel of Figure~\ref{fig:lvlcomp} when comparing our HST H$\alpha$+\Nii{} image to that of LVL. We also included uncertainties when multiplying the HST H$\alpha$+\Nii{} image by the \Nii/H$\alpha$ ratio derived from the pDSLM data (see Section~\ref{sec:methods}). The uncertainty in the pDSLM \Nii{} and H$\alpha$ images was again defined by measuring the MAD in emission-free regions, but additional errors from the flux calibration were included, as described in \cite{Lokhorst2022}. Once the intensity error map was created following this procedure, we took the average uncertainty in each of the cloud structure regions and attributed it to the corresponding mean intensity listed in column 9 of Table~\ref{table:values}.

We note, however, that there are still various unknowns in the parameters used to derive the densities, masses, and timescales shown in Table~\ref{table:values}. Thus, these errors should be considered underestimates. As shown for cloud 12 in Section~\ref{sec:cloud_phys_props}, the geometric uncertainties can cause $n_{\rm e}$ to vary by a factor of 3, $N_{\rm H}$ by a factor of 3/4, and $t_{\rm cc}$ by a factor of 1/3. In the case of $\chi$, the values rely on the density measurements of the hot wind from \cite{Lopez2020}, which did not give statistical uncertainties from the analysis or systematic uncertainties associated with the assumed volume filling factor and geometry. Works focusing on diffuse X-ray measurements of starburst winds \citep{Lopez2023} generally find statistical uncertainties in the derived densities of $15-50$\%. However, these measurements depend on the volume filling factor of the hot gas ($f_{\rm hot}$) as $n_{\rm e} \propto 1/\sqrt{f_{\rm hot}}$, and they assume $f_{\rm hot}=1$. Whether the hot gas fills the entire volume remains a matter of debate, depending on whether the observed soft ($<2$~keV) X-ray emission originates from the very-hot ($10^7$~K) wind fluid that has cooled or from interactions between the wind fluid and cooler material, as suggested by the presence of charge exchange \citep{Okon2024}. \edit1{Similarly, $f$ is unknown for these H$\alpha$ emitting structures and would affect the derived $n_{\rm{e}}$ by a factor of $1/\sqrt{f}$ (eq.~\ref{eq:ne2}). Thus, these geometric uncertainties indicate our $n_{\rm{e}}$ values are lower limits. This filling factor also propagates to the $N_H$ and $M_{cl}$ as a multiplicative factor of $f^{1/2}$ which means that by assuming $f=1$ the column densities and masses are upper limits.}
Lastly, $t_{\rm cc}$ also depends on the hot wind speed, which is unknown. Estimates range from $\sim1000$ $-$ $\sim2000$~km/s, introducing an additional uncertainty in $t_{\rm cc}$ of about a factor of 1/2.

\subsection{Multi-wavelength brightness profiles of the observed structures}
\label{sec:mutliwave_prof}

\begin{figure*}[ht!]
    \centering
    \includegraphics[width=0.94\textwidth]{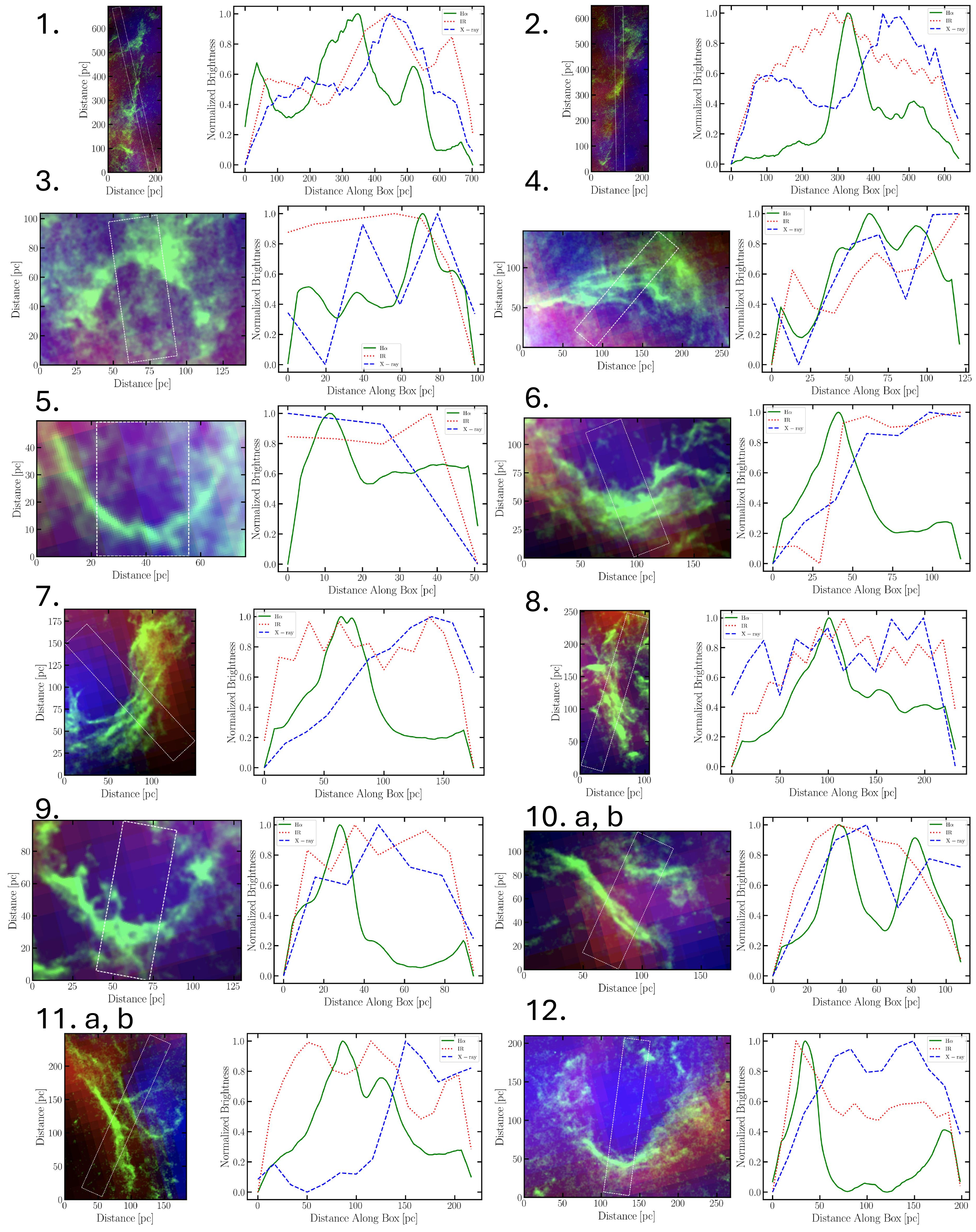}
    \caption{Three-color images each of the cloud regions where red is Spitzer 8~$\mu$m infrared, green is HST H$\alpha$ emission, and blue is \textit{Chandra} 0.5-7 keV X-ray emission. Right of each image are normalized brightness profiles showing where one wavelength dominates over the others across the image. The profiles are from bottom to top of the white boxes and are created by summing along its shorter side. }
    \label{fig:profiles}
\end{figure*}

To better understand the interplay of the ISM phases in the wind, we use the X-ray and infrared data described in Section~\ref{sec:methods} and compare their locations with the H$\alpha$. In Figure~\ref{fig:profiles}, we show three-color images for each of the regions and their normalized brightness profiles along the white box in the three-color image for each wavelength. The profiles are from bottom to top of the white boxes and are created by summing along its shorter side. For the first four regions the starburst is south of the box and for the others, the starburst is toward the north. The three-color images show X-ray-emitting gas \citep{Lopez2020} in blue, H$\alpha$ in green \citep{Mutchler2007}, and Spitzer 8~$\mu$m \citep{Kennicutt2003,Engelbracht2006} in red.

A few relationships are worth noting when inspecting the images and their corresponding profiles. We find an offset between the peaks of the X-ray and H$\alpha$ emitting structures in regions 1, 2, 3, 6, 7, 9, 10, 11, 12, where the X-ray decreases in intensity when it encounters the H$\alpha$. This behavior of the H$\alpha$ leading the X-ray has long been noted in the literature \cite{shopbell_asymmetric_1998,Lehnert1999,Stevens2003,Strickland2004} and signals the interaction of the hot wind with the cooler surrounding material that is being ionized. However, typically the correspondence between the H$\alpha$ and X-ray is noted on large kpc scales. Here we find that even on smaller $\sim100$~pc scales in the outflow the same behavior occurs. 

In the case of regions 4, 5, and 8, the relationship between the H$\alpha$ and X-ray emission is not as clear. In region 4 the two wavelengths seem to overlap significantly, and in region 8 the X-ray remains constant throughout the region. These clouds may be physically more porous than the other structures, allowing hot X-ray emitting gas to move through the cooler material, or projection effects may complicate the interpretation because hot gas may fill the region in front of the highlighted cool structures. 

Less clear in the profiles of Figure~\ref{fig:profiles} is the role of the infrared emission tracing the dust. In regions 11 and 12, the H$\alpha$ and infrared emission correlate in the images and brightness profiles. It may be the case that both wavelengths are tracing the same gas structure but at different locations. Since the dust is mixed with cooler neutral gas, it may sit at the center of the structure, whereas the outer layer is ionized, producing the H$\alpha$ emission we observe. This kind of correlation is also found in both \cite{Bolatto2024} and \cite{Fisher2024}, where high resolution ($\sim2$~pc) JWST images near the base of M82's outflow show this relationship between Pa$\alpha$ emitting structures and PAH emission. However, the other profiles do not show such behavior and would benefit from future work and observations with higher resolution instruments like JWST. 


\section{Discussion}\label{sec:discuss}

\subsection{Cloud Morphology and Possible Origin}

\label{sec:clmorphology}

An important property of the H$\alpha$ emitting structures we highlight in this paper is their non-cometary morphology, especially in the southern outflow. In the H$\alpha$ image, even the clouds approximated as ellipses in our analysis -- despite their long heights and narrow widths -- do not display a comet head at either end. Even more puzzling is the observation of arc-like clouds (e.g., Regions 3, 4, 5, 6, 7, 9, 11b, and 12 in Figure~\ref{fig:cloudatlas}) whose morphologies contrast with the cometary structures seen in simulations (e.g., \citealt{Cooper2008,Schneider2020}). These arc-like features resemble those identified in high-resolution JWST observations of PAH emission near M82's base by \cite{Bolatto2024}. However, as \cite{Fisher2024} highlighted, comet-like structures are also present in the same region. In contrast, we do not observe such morphologies at the larger distances ($\gtrsim 0.5$~kpc) analyzed in this study. 

The origin of the arc-like clouds is unclear but may be related to instabilities in the hot wind or to interactions between the hot wind and the cool pre-existing halo material from the tidal interaction with M81 or from the cool material that sheaths the M82 bicone. \cite{Yun1994} and \cite{deBlok2018} mapped the atomic hydrogen surrounding M82 using the Very Large Array (VLA) and identified large-scale tidal structures caused by interactions with its neighbor, M81. The wind may thus be interacting with the cooler halo gas, producing the observed H$\alpha$ emission. Another alternative is that the faster components of the wind are interacting with the slower neutral HI in the wind itself mapped by \cite{Martini2018}.

In particular, we note the strong morphological similarity with the nebula M1-67 around the Wolf-Rayet star WR 124 imaged with HST by \cite{Grosdidier1998}. Arc-like shapes are prevalent at the limb of the nebula and have been analyzed by \cite{Grosdidier2001}. An interpretation of the arc-like clouds in M1-67 suggested by \cite{Grosdidier1998} is that they arise from shocks generated as faster-moving clouds interact with a slower surrounding medium. The numerical models of \cite{Stone1995} indicate that for particular density profiles of the surrounding medium and time evolution of the wind power (specifically, if internal shocks are generated because a fast outflow is preceded by a slower wind phase), the wind-driven shell can become Rayleigh-Taylor and thin-shell unstable, producing knots of denser gas that plow into the slower moving medium. Under this hypothesis, the arcs seen in the HST images presented here are the shocks of faster-moving dense structures running into and decelerating in slower-moving surrounding gas, producing a bow shock-like morphology. The difference in the cloud shapes between the northern and southern M82 outflows might then be partially explained by the difference in the medium the M82 wind has expanded into in each hemisphere. Such a picture of the flow makes predictions for high-resolution IFU observations, which might reveal that the optical line-emitting structures are decelerating on large scales, particularly to the south. Further, the shape and size of the shocks may directly inform our picture of the interaction, the instabilities at work, and the time history of the wind power of M82.

Such a picture motivates new numerical simulations. For example, simulations analogous to \cite{Cooper2008} and \cite{Schneider2020,Schneider2024} generate cool filamentary and cometary structures, whose acceleration is dictated by momentum and energy exchange with the hot medium and cloud growth/survival criteria (e.g., \citealt{gronke_survival_2022}), but the arc-like structures we observe may be generated as the surviving population of clouds seen in these simulations interacts with a dense surrounding halo medium as in M1-67, motivated by the tidal structure observed in the M81-M82 system. Such a hypothesis could be directly tested by a series of global numerical simulations of winds interacting with a denser near-galactic or circumgalactic halo medium.

The structures may be further affected by the bicone shape of the hot outflow. In the standard picture, the hot ($10^{6-7}$~K) X-ray emitting gas occupies the core of a bicone, and it is surrounded by a ``sheath" of cooler material. The velocity difference between the faster hot gas and the slower-moving cooler phases at the bicone boundary will create shear, which may trigger Kelvin-Helmholtz instabilities and generate bow shocks that appear from our perspective as arcs. 
 
As a more speculative alternative, the arcs we see could potentially be created by instabilities associated with cosmic rays (CRs), which are known to be prevalent in the M82 starburst outflow \citep{Seaquist1991,Adebahr2013,Adebahr2017,Buckman2020} and have been suggested as a driver of cool gas in galactic winds (\citealt{Ipavich1975}; for a review, see \citealt{Ruszkowski2023,Thompson2024}). For recent global simulations with CRs, see \cite{Thomas2024}. CR-driven winds mediated by CR streaming transport are known to be unstable, producing shocks and shock trains and a thermally unstable medium \citep{Huang2022,Quataert2022,Tsung2022,Modak2023}. Additionally, the physics of individual cloud acceleration by streaming CRs in a complicated background medium may generate the arc-like, two-tailed morphology \cite{Bruggen2020}. It is interesting to speculate that the observed arcs may constrain or provide evidence of the large-scale physics of wind driving. A clue may be that the physical size of the features appears roughly constant with distance from the starburst, though the pressure of the hot outflow must decrease, even in a more collimated geometry.

\subsection{The Origin of the H$\alpha$ Emission}
\label{sec:ha_origin}

As mentioned in Section~\ref{sec:results}, the three-color profiles shown in Figure~\ref{fig:profiles} illustrate where the different phases coincide and differ in their brightness peaks. We find that for the majority of regions we analyze, the H$\alpha$ emission peak is offset from that of the X-ray, with the latter emission decreasing as if the hot wind encounters an obstacle. The spatial relationship between X-rays and H$\alpha$ has long been observed in M82 \citep{shopbell_asymmetric_1998,Lehnert1999,Stevens2003,Strickland2004} where the origin of the latter is typically attributed to interactions between the hot and cold phases of the wind. Recent work by \cite{Okon2024} strengthens this connection by analyzing the spatial distribution of charge-exchange (CX) emission in M82's hot wind using XMM-Newton data. CX occurs when an ion captures an electron from a neutral atom, emitting soft X-rays in the process, and in the context of galactic winds, it marks regions of interaction between the hot wind and cooler material. \cite{Okon2024} found that areas of enhanced CX emission coincide with H$\alpha$ filaments, reinforcing the link between these phases. These findings, along with the profiles in Figure~\ref{fig:profiles}, suggest that the hot wind, at least partly, plays an active role in shaping and ionizing the H$\alpha$-emitting structures.

The hot wind can ionize the cooler material through shocks, and past work indicates the process occurs in M82's outflow. \cite{Lehnert1996} studied the ionized gas in the halos of 55 edge-on starburst galaxies using long-slit spectroscopy across the major and minor axes of their targets. They find that about half their sample, including M82, have ``shock-like" line ratios of \Sii/H$\alpha$ and \Oi/H$\alpha$ along their minor axes. \cite{shopbell_asymmetric_1998} studied the warm gas in M82's outflow using Fabry-Perot data that include H$\alpha$, \Nii{}, and \Oiii{} spectral lines. Their work constrained the kinematics of the warm phase and used line ratios to infer the ionization mechanisms. They found that the \Oiii/H$\alpha$ ratio increases with distance from the starburst core. Combined with the spatial correspondence between H$\alpha$ and soft X-ray emission, they concluded that shock ionization dominates at large radii ($>$ 1~kpc). Within 1~kpc, they attributed the ionization to a combination of photoionization and shock ionization. \cite{Beirao2015} mapped the northern and southern outflow of M82 using the \textit{Spitzer}-IRS. They found that the warm molecular gas to PAH ratio is not consistent with ionization of atomic gas and is instead explained by shocks. 

Based on the literature, it is likely the H$\alpha$ emission we observe in these structures are a result of shocks caused by the hot wind at least at $>$ 1~kpc distances. Closer to the starburst, photoionization can still play a role as mentioned in \cite{shopbell_asymmetric_1998} and in \cite{Westmoquette2009b} who found that for the inner 0.9~kpc of M82, the line ratios can be explained by pure photoionization. Thus in our analysis, it is possible that the emission from clouds 4, 5, 6, and 7 can be due to photoionization; however their arc-like nature makes it difficult to rule out shocks completely. 


However, an important issue is if the H$\alpha$ emission is a result of either shocks or photoionization, it may be that we are only observing the outer layers of the cloud where the cool gas meets the hot wind. If the dust traces structures similar to the ionized gas (which only appears on their surface), this could suggest that the clouds have even cooler atomic, and possibly molecular, gas in their cores. In that case, the H$\alpha$ may represent the turbulent mixing layers modeled in simulations that can then cool and even grow the clouds \citep{fielding_structure_2022}. Supporting this hypothesis, \cite{Fisher2024} found a correlation between the plumes they identified near M82's base in the infrared and Pa$\alpha$ emission. 

\subsection{Possible Cool Cloud Acceleration Mechanism}
\label{sec:cloud_accel}

Although the mechanism responsible for cool gas acceleration is not known, a useful benchmark for acceleration comes from comparing our derived values for the cloud column densities to the maximum cloud column that could be accelerated by a hot outflow via ram pressure with force $\dot{p}_{\rm hot}=\dot{M}_{\rm hot}v_{\rm hot}$. For a gravitational acceleration of the form $g = 2\sigma^2/r$, equating the hot wind ram pressure with the gravitational force yields a maximum critical cloud column density projected along the wind axis of $N_{\mathrm{cl,Edd}} \simeq (\xi/4)(\dot{p}_{\rm hot}/(\Omega \sigma^2 r))$ (the ``Eddington" column in \citealt{Thompson2024}). Writing $\dot{p}_{\rm hot}$ in terms of the thermalization efficiency $\alpha$ and the mass-loading of the hot outflow relative to the star formation rate ($\eta=\dot{M}_{\rm hot}/{\rm SFR}$), the maximum column can be written as
\begin{equation}
    N_{\mathrm{cl,\,Edd}} 
    \simeq8\times10^{21}\;\mathrm{cm}^{-2}\frac{\xi(\alpha \eta)^{1/2}\mathrm{SFR_{10}}}{R_{0.3}\Omega_{4\pi}\sigma_{100}^2},
    \label{eq:eddcol}
\end{equation}
where $\xi$ is the cloud drag coefficient, ${\rm SFR}_{10}={\rm SFR}/10$\,M$_\odot$ yr$^{-1}$, $R_{0.3}=R/0.3$\,kpc is the starburst core radius, $\Omega_{4\pi}=\Omega/4\pi$ is the solid angle subtended by the hot wind, and $\sigma_{100}=\sigma/100$\,km s$^{-1}$. Clouds with projected column densities below this value can be accelerated by the momentum injection of the hot flow, while clouds with column densities above this critical value cannot.


Several of the parameters in Equation~\ref{eq:eddcol} are constrained by observations. In particular, the X-ray morphology of nearby starburst winds \citep{Lopez2020,Lopez2023,Porraz2024} and theoretical models \citep{Nguyen2021} indicate that 
X-ray emitting gas is confined to a bicone and more cylindrical than spherical, affecting $\Omega$ for the hot wind. For M82, using an opening angle for the hot outflow of $\theta\simeq62^{\circ}$ \citep{Strickland1990},
$\Omega_{4\pi} \simeq 0.14$.
Further, \cite{Greco2012} 
found a rotational velocity of $v_{\rm rot}\sim120$~km~s$^{-1}$ at a disk radius of 1~kpc,
implying $\sigma\simeq85$\,km s$^{-1}$.
Additionally, \cite{Strickland2009} used hard X-ray data in the core of M82 to constrain 
$0.3 \leq \alpha \leq 1$ and $0.2 \leq \eta \leq 0.6$, implying that $(\alpha\eta)\simeq0.06-0.6$.
Assuming $\xi\simeq0.5$, $\Omega\simeq0.14$, and using this range in $(\alpha\eta)$, we find that $N_{\mathrm{cl,Edd}}\simeq1-3\times10^{22}$\,cm$^{-2}$ on a scale of $R=0.3$\,kpc. On larger scales, the critical column for acceleration decreases with distance so that at $\simeq3$\,kpc, $N_{\mathrm{cl,Edd}}$ would be ten times smaller: $N_{\mathrm{cl,Edd}}\simeq1-3\times10^{21}$\,cm$^{-2}$ at $R=3$\,kpc. While the uncertainties in $N_{\mathrm{cl,Edd}}$ preclude a precise comparison, the distances ($0.5-2.6$\,kpc) and column densities ($2-14\times10^{21}$\,cm$^{-2}$) in Table \ref{table:values} indicate that gravity may play an important role in the dynamics of the structures we highlight here. For an analogous discussion of the potential importance of radiation pressure on dust on kpc scales in the outflow of M82, see \cite{Coker2013}.

\subsection{Comparison with Simulations}
\label{sec:comp_sims}
Recent simulations have made significant progress in modeling large-scale multiphase winds \citep{Schneider2020,fielding_structure_2022} as well as in conducting simulations that focus on cloud survival and the interface between the cool and hot gas \citep{gronke_growth_2018,Gronke2020,gronke_survival_2022,Abruzzo2023,Tan2024,Villares2024}. Even though the structures we observe do not look like the ones in these cloud simulations, we can still make several useful comparisons that are detailed below. 

One of the primary quantities that theoretical works have used to gauge whether or not clouds can survive in the hot wind is a minimum cloud radius. \cite{gronke_growth_2018} derived a criterion for this radius based on the wind Mach number $M_{\rm wind}$, cloud temperature, thermal pressure, and density contrast between the cool clouds and the hot wind:
\begin{equation}
    r_{\rm{crit, cc}} = 2\:\mathrm{pc}\:\frac{T_{\rm 
cl,4}^{5/2}\:M_{\rm wind}}{P_3\:\Lambda_{\rm mix,-21.4}}\frac{\chi}{100}\:f(M_{\rm{turb}}),
    \label{eq:cloud_crit}
\end{equation}
where $T_{\rm cl,4} \equiv (T_{\rm cl}/10^4\mathrm{~K})$, $P_3 \equiv nT/(10^3\mathrm{~cm}^{-3}\mathrm{~K})$, and $\Lambda_{\rm mix,-21.4} \equiv \Lambda(T_{\rm mix})/(10^{-21.4}\mathrm{~erg~cm}^3\mathrm{~s}^{-1})$. The function $f(M_{\rm{turb}})$ is from \cite{gronke_survival_2022}, which has the empirical relation $f(M_{\rm{turb}})\sim 10^{0.6M_{\rm{turb}}}$ to account for the wind's turbulent disruption of the cloud.  Equation~\ref{eq:cloud_crit} was later modified by \cite{Abruzzo2023} based on the work of \cite{Li2020} and \cite{Sparre2020} to incorporate the effects of shear from the wind on the cloud. Essentially, as the hot wind engulfs the cloud, the interface between the two phases must cool on a shorter timescale than the wind crossing time; otherwise, the material will be advected by the wind, preventing cloud growth. Thus, they argue that the final cloud radius criterion is given by,
\begin{equation}
    r_{\mathrm{crit,shear}} = \sqrt{\chi}\;r_{\mathrm{crit,cc}}.
    \label{eq:cloud_shear}
\end{equation}
We find that the structures we observe are larger than the $r_{\rm crit, shear}$ derived in several theoretical works. Using the criterion in Equation~\ref{eq:cloud_shear}, \cite{Tan2024} found a minimum cloud radius of approximately 5~pc for their fiducial simulation and cloud radii between 8 and 100~pc. The models of \cite{fielding_structure_2022}, find inferred cloud radii (calculated by \citealt{Xu2023}) that range between approximately 5 and 200~pc. Our measurements for cloud radii ($r_{\rm cl}$) range from about 14 to 110 pc. These measurements are well within the range of these simulations and exceed the minimum cloud radius they derive.

Another way to write the minimum radius for cool cloud survival in a hot wind is as a minimum column density, above which clouds would be expected to survive and grow. \cite{Thompson2024} writes this minimum column density for growth from the criterion of \cite{gronke_growth_2018} as 
\begin{equation}
    N_{\mathrm{cl,grow}} \simeq 5\times10^{18}\;\mathrm{cm^{-2}}\left(\frac{\alpha}{\eta}\right)^{1/2}T_{\mathrm{mix},5.5}\;\Lambda_{-21.4},
    \label{eq:col_crit}
\end{equation}
where $\alpha$ and $\eta$ are the wind thermalization and mass-loading efficiencies, $T_{\mathrm{mix},5.5}$ is the temperature of the mixed gas at the cloud surface in units of $10^{5.5}$~K, and $\Lambda_{-21.4}$ is the cooling function for $T_{\mathrm{mix}}$ in units of $10^{-21.4}\;\mathrm{ergs\;cm^3\;s^{-1}}$. We find column densities on the order of $N_{\rm H} \approx 10^{20} - 10^{21}\;\rm{cm^{-2}}$. These values are well above the criterion in Equation~\ref{eq:col_crit} assuming the nominal values \edit1{and $f=1$,} indicating that the structures we observe are likely to survive to distances of a few kpc. The simulations of \cite{fielding_structure_2022} also find column densities above the threshold, with values at a distance of 1~kpc ranging from about $1\times10^{19}\:\rm{cm^{-2}}$ to $8\times10^{20}\:\rm{cm^{-2}}$ (as reported in \citealt{Xu2023}).

We also find that the values of $\chi$ we derive are similar to simulations. \cite{gronke_survival_2022} show in their simulations studying cloud survival in a turbulent medium, rather than the laminar flows of earlier work \citep{gronke_growth_2018}, that runs with $\chi\sim100$ exhibit cloud survival,  whereas runs with $\chi\sim1000$ do not. The simulations of \cite{Tan2024} find this value to range from $10^2$ to $10^3$ with an average of about 400. Our values for $\chi$ in Table \ref{table:values} are in the range $28 - 240$, with the majority being of order $10^2$. Compared to simulations of \cite{gronke_survival_2022} with find agreement with our measurements. Compared to \cite{Tan2024}, our values are similar, but on the lower side of their distribution.

When relating the cool cloud pressures ($P_{cl}$) to the hot wind pressures ($P_X$), we find agreement with the simulations of both \cite{Tan2024} and \cite{Schneider2020}. They find that most of the cool clouds are not in thermal pressure equilibrium with the hot phase: the cool clouds have smaller pressures than the hot phase, which, as asserted in \cite{Schneider2020}, may be a result of the cooling time of the cloud being shorter than the sound crossing time. Using our derived gas densities and assuming $T = 10^4$~K, we find $P_{cl}/k\sim8.4\times10^4\:-\:2.3\times10^5\mathrm{~cm}^{-3}\mathrm{~K}$, which, when compared to the pressures of the hot phase from \cite{Lopez2020} that range from $P_{\rm X}/k = 10^5-10^7\;\mathrm{~cm}^{-3}\mathrm{~K}$, are underpressurized.

We can also compare the relevant timescales of entrained cool clouds in M82 using the values in Table~\ref{table:values}. As mentioned earlier, \cite{Abruzzo2023} argue that not only should the cooling time for the mixing layer of the cloud be less than the cloud-crushing time, but also that of the cloud shear timescale, defined as $t_{\mathrm{shear}} = r_{\rm cl}/v_{\rm w}$. For cloud survival, they find that $t_{\mathrm{shear}}\gtrsim7t_{\rm cool, mix}$. Assuming $v_{\rm w}=2000$~km~s$^{-1}$, we find $t_{\mathrm{shear}} \approx 0.02$~Myr, and with a factor of 7 from the models of \cite{Abruzzo2023}, $t_{\mathrm{shear}}$ increases to about 0.16~Myr. We note that this is a lower limit, as we assume the structures' velocities are zero; without knowing their true values, we cannot accurately determine the relative velocity required to calculate $t_{\mathrm{shear}}$. Assuming $T_{\mathrm{mix}} = 10^{5.5}\:{\rm K}$ and $\Lambda = 10^{-21.4}\;\mathrm{ergs\;cm^3\;s^{-1}}$, we find the cooling times of our cloud mixing layers to be between  $4.7\times10^{-4}-1.3\times10^{-2}$~Myr. Thus, our cloud values are well below both the typical cloud crushing and cloud shear criteria, indicating they have the potential to grow as mass from the wind is transferred to the cloud.

In summary, the various parameters from simulations studying cloud survival such as $\chi$, $N_{\rm H}$, $r_{cl}$, pressures, and $t_{\rm cool}$, are consistent with our findings. However, as discussed in Section~\ref{sec:clmorphology}, the major discrepancy between theory and observations in this work is the non-cometary morphology of the clouds, which is a persistent and conspicuous feature of (especially) the southern outflow. Future simulations may aim to study how these diverse morphologies can form while maintaining the agreement with the other parameters mentioned.


\subsection{Comparison with previous observational work}
\label{sec:comp_obs}
While much work has been done on M82's multi-phase galactic wind, there has been little work attempting to physically resolve and directly constrain the properties of small-scale structures swept up by the wind. However, in recent years, constraints have been made in molecular gas \citep{Krieger2021}, optical absorption line \citep{Xu2023}, and infrared studies \citep{Fisher2024} that we can compare with our values in this paper.

We find that our radius and mass measurements are consistent with the constraints in the molecular phases from \cite{Krieger2021}. Their work analyzed CO (1-0) observations from NOEMA and resolved molecular gas structures in M82 for several regions, though we will only focus on the northern and southern outflow regions. They used the FellWalker algorithm \citep{Berry2015}, which employs a clump finder and watershed algorithm. With a resolution of about 30~pc, \cite{Krieger2021} found median cloud radii ranging from approximately 30$-$60~pc and masses ranging from $10^{4.5}-10^{6}\;\rm{M_\odot}$, though the exact range varies depending on the CO-to-H$_2$ factor used, and both quantities decreasing with distance from the nucleus. Our values for mass are within the range they find, and our radii are the same order of magnitude as theirs with some structures marginally below or above their range. 

When comparing with the optical absorption line studies of \cite{Xu2023}, we do not find agreement with our constraints on $n_{\rm e}$ and $r_{\rm cl}$ but do find consistency with our $N_{\rm H}$ measurements. \cite{Xu2023} used H$\alpha$ images from \cite{Ohyama2002} and spectra from \cite{Yoshida2019} to constrain the sizes and column densities of clouds in the ionized optical-emitting phase of M82's wind. \cite{Yoshida2019} calculated the number densities of the ionized gas using the [SII] $\lambda\lambda6717,\:6731$ doublet, which \cite{Xu2023} rebinned to minimize scatter and derive a power-law scaling with distance from the disk. \cite{Xu2023} found a decrease in number density that scales as $n_{\rm e}(r)=100\times(r/1165\:\rm{pc})^{-1.17}\:\rm{cm^{-3}}$, yielding a number density of about $120\:\rm{cm^{-3}}$ at 1~kpc, 10$-$100 times the densities we derive from the individual structures in Table~\ref{table:values}. 

To constrain cloud radii, \cite{Xu2023} calculated a filling factor for the ionized gas by using their number density constraints \edit1{from the \Sii{} doublet} and assuming a conical outflow shape to estimate the total volume. \edit1{Similar to Eqn.~\ref{eq:ne2}, they divide the H$\alpha$ surface brightness by $n_e^2$, the H$\alpha$ energy and recombination coefficient, and the path length through the outflow that they measure to be 1.5$r$ where $r$ is the vertical distance to the starburst center.} Then, using the same path length through the outflow as well as a covering factor derived from the median of covering factors from the CLASSY sample of low-redshift star-forming galaxies \citep{CLASSY}, they found cloud radii in the range 0.07$-$0.9~pc, two orders of magnitude smaller than our constraints, those of \cite{Krieger2021}, and below the resolution of HST imaging. \cite{Xu2023} also calculated the column density of the clouds using their radius constraints and number densities, yielding column densities in the range of $10^{19.1}-10^{20.7}\:\rm{cm^{-2}}$. These values are similar to those of the theoretical work of \cite{fielding_structure_2022}, meet the minimum cloud column requirement of Equation~\ref{eq:col_crit}, and are similar to our lowest column density constraints. 

\edit1{The question then arises of why the $n_{\rm e}$ values between this work and \cite{Xu2023} are significantly different. The answer to this lies in the filling factor. If we take the same regions as \cite{Xu2023} along the outflow, use our intensity values, assume the same path length and $n_e$ they measure, and then solve for $f$, we find values of order $10^{-4}$ consistent with \cite{Xu2023}. This would indicate that the H$\alpha$ emission we observe pertains solely to a limited volume, likely on the cloud surface (see Section~\ref{sec:ha_origin}). 
An alternative explanation that does not require a filling factor is that the $\Delta s$ we assume is substantially smaller than assumed. However, in order to reconcile the two densities, the clouds would require a sheet-like geometry with a depth into the sky of $\sim0.1$\,pc, much smaller than the width observed in the plane of the sky. As discussed in the previous sections, the known shock-like line ratios of the outflow past 1~kpc along with the multiwavelength profiles in Figure~\ref{fig:profiles}, leads us to surmise that the first hypothesis is more likely. If $f$ is as small as reported in \cite{Xu2023} for the structures we observe, then our reported densities and column densities in Table 1 are substantial underestimates and overestimates, respectively, given equations \ref{eq:ne2} and \ref{eq:col}.
}

\edit1{However, it is worth noting that if the filling factor is as low as predicted in \cite{Xu2023} -- which scales as $f=10^{-3}\times(r/628\;\mathrm{pc})^{-1.8}$ where $r$ is the radial distance form the starburst-- the cloud column density becomes problematic for cloud survival. Equation~\ref{eq:col_crit} gives an estimate for  the lowest column density allowed for a cloud to survive and grow in the hot wind. Assuming $\alpha=1$ and $\eta=0.6$ (as derived in \citealt{Strickland2009}), the minimum cloud column density becomes $N_{\rm{cl,grow}}=6.5\times10^{18}\;\mathrm{cm^{-2}}$. At the location of cloud 12 (1.8~kpc), $f=1.5\times10^{-4}$ and would result in $N_{\rm{cl}}=2.3\times10^{18}\;\mathrm{cm^{-2}}$, below the survival criterion.
However it is also unlikely that the filling factor is unity. \cite{Westmoquette2009b} also found small filling factors in M82's wind that are the same range as \cite{Xu2023} and imply sub-parsec cloud sizes. In another analogous starburst, NGC~253, \cite{Cronin2025} find clumping factors of about $0.1-0.4$, noting that too high of a clumping factor (i.e. uniform density), would result in exceeding the energy budget for the system. This also leads to the conclusion, that while our measured cloud mass upper limits are not extraordinarily high ($\sim10^4\;\rm{M_\odot}$), extrapolating to larger regions the size of those in \cite{Xu2023} could potentially result in the warm mass (calculated as $M_{cl}=nf^{1/2}V$) exceeding the energy budget constrained by the hot phase results of \cite{Strickland2009}.}

\edit1{Finally, we note that we} find good agreement with the size of structures found in \cite{Fisher2024} but not their morphology. They used 3.3~$\mu$m infrared JWST observations near the base of M82's wind to characterize various filamentary structures. They identified four plumes in each outflow hemisphere, defining them as coherent flux peaks extending roughly perpendicular from the disk plane. The width of each plume was measured to be approximately 30 to 40~pc, with heights extending to the edge of their observation window, corresponding to 300$-$400~pc in height. These plumes are similar in size to clouds whose shapes were approximated as ellipses in Table~\ref{table:values}. The identified plumes were further decomposed into 12 individual clouds with widths of 5$-$18~pc and heights of 30$-$150~pc \cite{Fisher2024}. These clouds are roughly consistent with our measurements, but their morphologies differ. Several of the clouds shown in \cite{Fisher2024} appear to have the cometary structures reported in simulations, where density peaks at the heads of the comets (clouds). In our work, besides the arc-like structures, our elongated clouds do not resemble comets rather they have varying density throughout the structure that does not peak at either end.

\section{Conclusions}

In this paper, we analyze H$\alpha$ emission from the outflow of M82 captured by \citet{Mutchler2007} using the F658N filter on HST, which we continuum subtracted and corrected for \Nii{} to isolate the nebular H$\alpha$ emission (Figure~\ref{fig:cont_sub}). We characterize the cool gas structures carried by the starburst-driven wind. For each cloud, we assign various geometric shapes to constrain number density, column density, and cloud radii shown in Table~\ref{table:values}. These cloud properties are then compared to simulations (Section~\ref{sec:comp_sims}), past observational work (Section~\ref{sec:comp_obs}), and various cloud survival criteria (Section~\ref{sec:cloud_accel}). We also discuss the diverse morphologies (Section~\ref{sec:clmorphology}) of the clouds, their possible origins (Section~\ref{sec:ha_origin}) and compare to multiwavelength data (Section~\ref{sec:mutliwave_prof}).  We make the following conclusions:

\begin{itemize}

	\item \textit{Morphology:} We observe a diverse set of cool-cloud morphologies in the wind (see Figures~\ref{fig:south_zoomin_usm} and \ref{fig:north_zoomin_usm}). We categorize them broadly into either elongated elliptical clouds or arc-like clouds. Neither set of clouds resembles the cometary structures predicted by simulations. We discuss the possibility that the arc-like clouds are bow shock-like structures formed by interactions with the surrounding HI-rich medium or, more speculatively, the result of cosmic ray-driven instabilities in the outflow (see Section~\ref{sec:clmorphology}). 
	
    \item \textit{Cloud Properties:} We compile the cloud properties in Table~\ref{table:values} and show their ranges in emission measure in Figure~\ref{fig:em_hists}. For the elliptical clouds, we find number densities of approximately $n_{\rm e} \approx 0.80-5.1$~$\rm cm^{-3}$, radii between $r_{\rm cl} \approx 24 - 82$~pc, and column densities of $N_{\rm H} \approx 4.2\times10^{20}-1.2\times10^{21}\;\rm{cm^{-2}}$. The arc-like clouds have number densities between $n_{\rm e} \approx 2.6-23\;\rm cm^{-3}$, radii of $r_{\rm cl} \approx 14.3-50$~pc, and column densities of $N_{\rm H} \approx 2.4\times10^{20}-1.2\times10^{21}\;\rm{cm^{-2}}$. We also estimate the mass of the clouds finding them to be $M_{\rm cl} \approx 6.9\times10^{3}-1.3\times10^{5}\;\rm{M_\odot}$ for the elliptical clouds and $M_{\rm cl} \approx 6.9\times10^{3}-1.1\times10^{5}\;\rm{M_\odot}$ for the arcs. \edit1{These constraints were made assuming a volume filling factor of unity ($f=1$), thus the density dependent values are lower limits (see Eqs.~\ref{eq:ne2} \& \ref{eq:col}).}

    \item \textit{Location of Mutliwavelength Emission:} We create multiwavelength brightness profiles and find that the peaks in H$\alpha$ are offset from peaks in the X-ray which subsequently drops as if it encounters an obstacle (see Figure~\ref{fig:profiles}). Recent work indicates that charge-exchange is occurring at the locations of these H$\alpha$ filaments, suggesting that the hot and cold wind phases are interacting and producing the observed optical emission. Along with past work studying the ionization structure of the outflow, our brightness profiles may indicate that shocks caused by the hot wind could be causing the optical emission in these cloud structures. We note, however, that we cannot rule out that  photoionization may be occurring near the base of the outflow, though it is unlikely to be the dominant mechanism due to the arc-like, bow-shock appearing cloud structures near the base.  

	\item \textit{Comparison to Simulations:} The observed cloud properties are mostly in agreement with both cloud-crushing and larger-scale galactic wind simulations. We find broad agreement with cloud number density, column density, mass, thermal pressures, and density contrast between the cool cloud and hot wind (see Section~\ref{sec:comp_sims}). The only major difference we find is the prevalence of arc-like structures differing heavily from the cometary structures found in these theoretical works. 

    \item \textit{Cloud Survival:} Our analysis shows that the cool cloud structures we identify are likely to survive within the hot galactic wind. We find cooling times of order $10^{-3}$~Myr, shorter than both the cloud-crushing timescale ($\sim10^{-1}$~Myr) and cloud shearing timescale ($\sim10^{-1}$~Myr) assuming a wind velocity of 2000~km s$^{-1}$. The observed clouds also lie between the minimum cloud column density set by the cooling time of the cloud mixing layer and the maximum column density set by the maximum ram pressure the wind can exert (see Sections~\ref{sec:cloud_accel} and \ref{sec:comp_sims}). Through these various metrics, we find that the observed structures not only survive but may even grown via mass exchange from the hot wind.

\end{itemize}

Future work will aim to utilize high-resolution optical and infrared spectroscopy of these structures to ascertain their properties with less bias from projection effects. Instruments like the Keck Cosmic Web Imager or the JWST NIRSpec Micro-Shutter Assembly can aid in revealing the true ionization mechanism of these clouds, as well as their kinematics, to more conclusively determine their origin and trajectory. Observations of M82 with XRISM~\citep{XRISM} will also allow for better constraints on the hot wind kinematics that can affect our derived cloud properties.

\acknowledgements
We thank Deborah Lokhorst and the rest of the pathfinder Dragonfly Spectral Line Mapper team for providing the narrow line data used in this analysis. We thank the OSU Galaxy/ISM Meeting for useful discussions and we thank Tim Heckman for a careful reading of the text and for  comments and suggestions that improved the manuscript. Support for this work was provided by the National Aeronautics and Space Administration through Chandra Award Number AR4-25007X issued by the Chandra X-ray Center, which is operated by the Smithsonian Astrophysical Observatory for and on behalf of the National Aeronautics Space Administration under contract NAS8-03060. SL and LAL were supported by NASA’s Astrophysics Data Analysis Program under grant No.
80NSSC22K0496SL, and LAL also acknowledges support through the Heising-Simons Foundation grant 2022-3533. TAT acknowledges earlier support by NASA 21-ASTRO21-0174. TAT thanks Drummond Fielding, Peng Oh, Greg Bryan, Eve Ostriker, Christoph Pfrommer, Timon Thomas, and Mateusz Ruszkowski for stimulating discussions at the Aspen Center for Physics, where part of this work germinated, and which is supported by NSF grant PHY-2210452. This work is based in part on observations made with the Spitzer Space Telescope, which was operated by the Jet Propulsion Laboratory, California Institute of Technology under a contract with NASA.


\vspace{5mm}

\software{astropy \citep{astropy,Astropy2018,Astropy2022}, scikit-image \citep{scikit-image}}

\bibliography{main}{}
\bibliographystyle{aasjournal}

\end{document}